\documentclass[lettersize,journal]{IEEEtran}
\usepackage{amsmath,amsfonts}
\usepackage{algorithmic}
\usepackage{algorithm}
\usepackage{array}
\usepackage{subcaption}
\usepackage{textcomp}
\usepackage{stfloats}
\usepackage{url}
\usepackage{verbatim}
\usepackage{graphicx}
\usepackage{cite}
\usepackage{color}
\usepackage{makecell}
\usepackage{pifont}
\newcommand{\cmark}{\ding{51}}%
\newcommand{\xmark}{\ding{55}}%

\newcommand{\figurewidthI}{0.7\columnwidth}
\newcommand{\figurewidthII}{0.9\columnwidth}
\newcommand{\reviewI}{\color{black}}
\newcommand{\reviewII}{\color{black}}
\newcommand{\reviewIII}{\color{black}}

\begin{document}


\title{Better Together: Leveraging Multiple Digital Twins for Deployment Optimization of Airborne Base Stations}

\author{Mauro Belgiovine,~\IEEEmembership{Student Member, IEEE},~Chris Dick,~\IEEEmembership{Senior Member, IEEE},~Kaushik Chowdhury,~\IEEEmembership{Fellow, IEEE}
}

\maketitle

\begin{abstract}
Airborne Base Stations (ABSs) allow for flexible geographical allocation of network resources with dynamically changing load as well as rapid deployment of alternate connectivity solutions during natural disasters. Since the radio infrastructure is carried by unmanned aerial vehicles (UAVs) with limited flight time, it is important to establish the best location for the ABS without exhaustive field trials. This paper proposes a digital twin (DT)-guided approach to achieve this goal through the following key contributions: (i) Implementation of an interactive software bridge between two open-source DTs such that the same scene is evaluated with high fidelity across NVIDIA’s Sionna and Aerial Omniverse Digital Twin (AODT), highlighting the unique features of each of these platforms for this allocation problem, (ii) Design of a back-propagation-based algorithm in Sionna for rapidly converging on the physical location of the UAVs, orientation of the antennas and transmit power to ensure efficient coverage across the swarm of the UAVs, and (iii) numerical evaluation in AODT for large network scenarios (50 UEs, 10 ABS) that identifies the environmental conditions in which there is agreement or divergence of performance results between these twins. Finally, (iv) we propose a resilience mechanism to provide consistent coverage to mission-critical devices and demonstrate a use case for bi-directional flow of information between the two DTs.

\end{abstract}


\section{Introduction}
\label{sec:introduction}

Unmanned Aerial Vehicle (UAV)-mounted Base Stations, or Airborne Base Stations (ABSs), have gained significant attention as a complement to ground-based cellular networks \cite{8918497}. As UAVs become more accessible, their ability to navigate {\reviewI 3-dimensional (3D)} space provides flexibility in adapting to dynamic network demands \cite{8713514, 7451189}, enabling line-of-sight links to mission-critical units \cite{en15155681} and enhancing user tracking \cite{8306379}. However, ABS-enabled connectivity introduces challenges such as collision avoidance, coordinated coverage, and optimal placement, considering limited flight times of 20 to 100 minutes~\cite{top10drones}. These challenges are highly dependent on the RF propagation environment, making prior channel knowledge essential for effective network planning.

\noindent$\bullet$ \textbf{Motivation for Digital Twins:} Optimal placement of Base Stations (BSs) is traditionally handled by telecom operators relying on domain knowledge and best practices. Various factors, including geography, network performance, and costs, influence these decisions. Digital Twins (DTs) and, specifically, Digital Twins for Networking (DTNs) \cite{10148936}, have emerged as strategic tools for network simulation, performance analysis, and ``what-if" scenarios. DTNs aid in planning, performance tuning, and machine learning-driven traffic modeling. In wireless networks, they also enable accurate propagation modeling, antenna design, and multi-antenna configurations, playing a key role in developing 6G systems and beyond.

\begin{figure}[t]
\centering
\includegraphics[width=0.95\columnwidth]{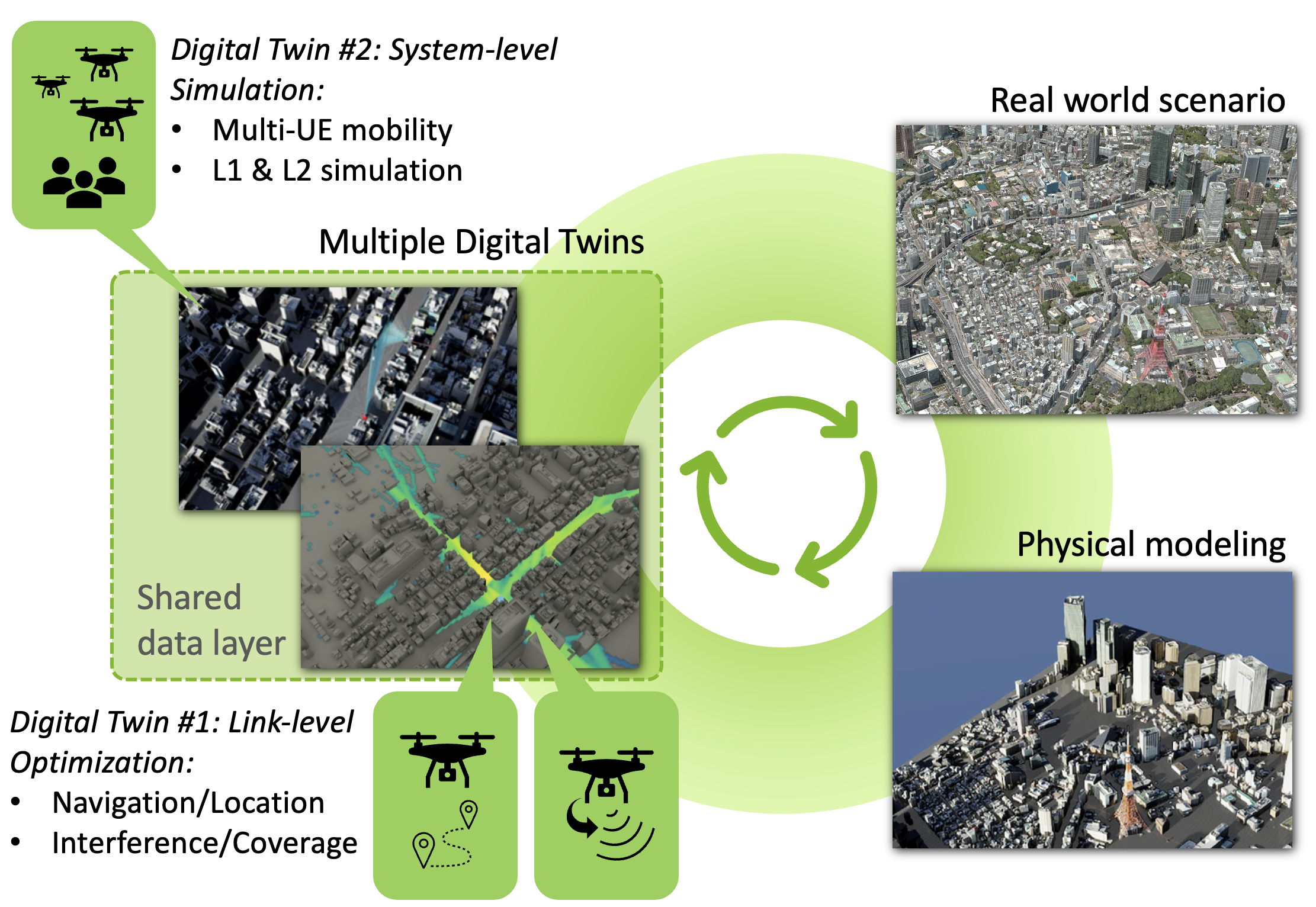}
\caption{Overview of proposed optimization and validation framework for Airborne Base Stations (ABSs) deployment using Multiple Digital Twins.}
\end{figure}

\noindent$\bullet$ \textbf{Challenges in Using Digital Twins:} Despite advancements in DTN tools, no single solution can comprehensively simulate complex wireless networks. Engineering such systems requires expertise in signal processing, propagation modeling, and software architectures. DTs offer varying capabilities, from network optimization to large-scale physical simulations. Integrating multiple DTs can enhance planning but introduces challenges such as 3D site model sharing, node placement consistency, and coherent interpretation of simulation results across different solvers.

\begin{figure}[t]
\centering
\includegraphics[width=0.95\columnwidth]{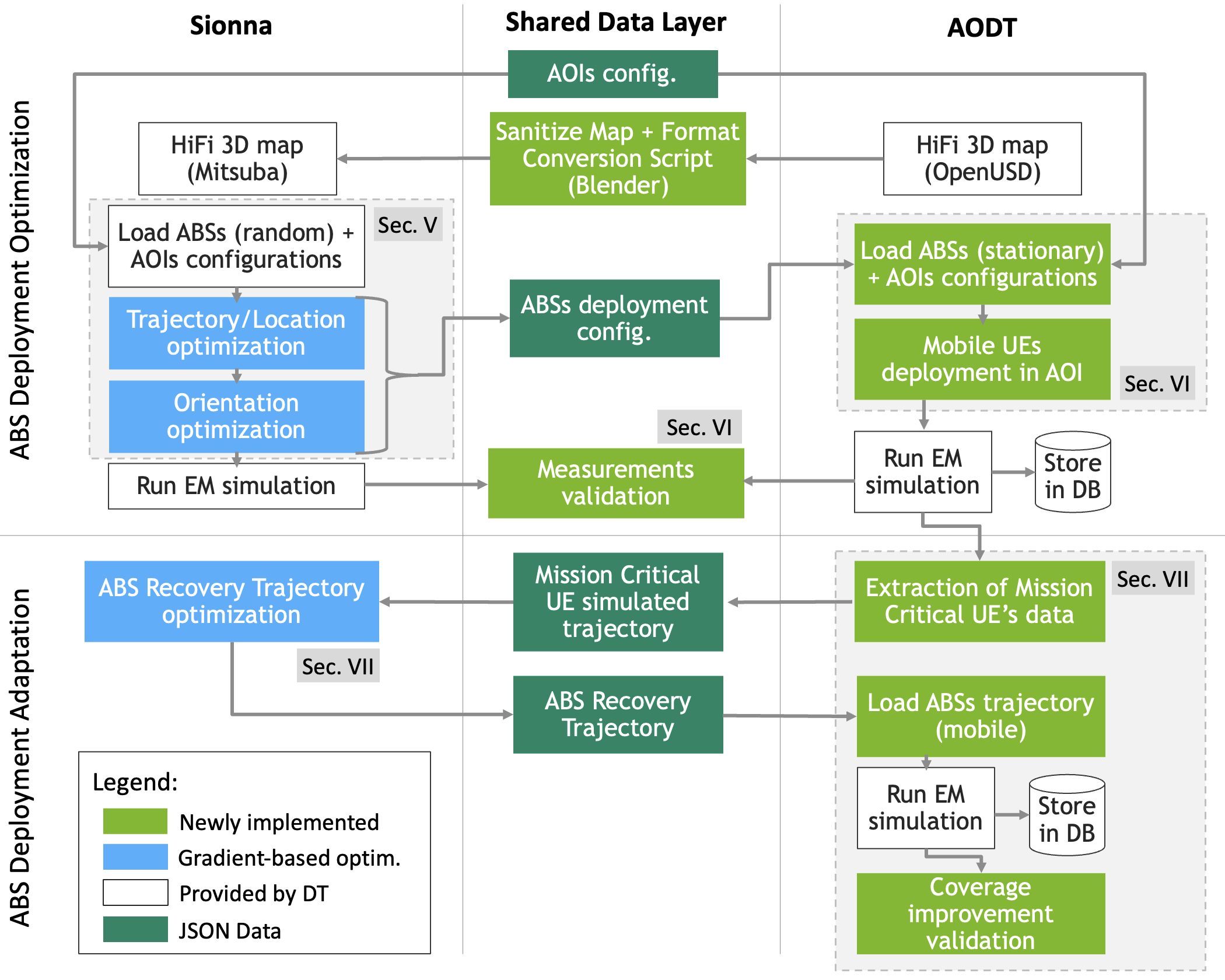}
\caption{{\reviewI  Multi-DT framework showing task separation: Sionna (left) performs gradient-based optimization, AODT (right) handles validation and mobility simulation, and the Shared Data Layer enables bidirectional communication through standardized data exchange (3D models, ABS configurations, User Equipment (UE) trajectories and simulation results). Arrows demonstrate the synergistic information flow between platforms at each computation step.}}

\label{fig:multiDT_implementation}
\end{figure}

\noindent$\bullet$ \textbf{Contributions of the Paper:} This work presents a Multiple-Digital Twin (Multi-DT) system for autonomous ABS deployment in city-scale environments. We integrate NVIDIA’s Sionna and Aerial Omniverse Digital Twin (AODT) to: (i) use Sionna’s differentiable simulation to optimize ABS trajectories and orientations, (ii) validate deployments with AODT-generated large-scale simulation data, (iii) leverage AODT data to enhance ABS resilience for mission-critical coverage, and (iv) bridge interoperability gaps between DTs to enable cooperative functionalities. This framework demonstrates the advantages of Multi-DTs for complex wireless tasks and promotes their adoption in research. Proposed implementation overview is shown in Fig. \ref{fig:multiDT_implementation}.

\section{Related Work}
\label{sec:relatedwork}

ABS deployments have been explored for various applications \cite{9681624, doi:10.1177/15501329221123933}, including enhancing network capacity in dense areas \cite{8713514}, supporting vehicular networks \cite{10.1016/j.vehcom.2022.100498}, and aiding disaster-affected regions \cite{9453842}. Autonomous ABS deployment is crucial in hazardous or inaccessible environments: {\reviewII Reinforcement Learning (RL)}-based approaches have been proposed for ABS deployment in coverage-limited areas, optimizing position and orientation for backhaul connectivity \cite{10056676, 9321160}. These models do not account for physical obstacles, limiting real-world applicability. {\reviewII Moreover, while RL approaches can be effective when ABSs have limited environmental data, they require extensive sensing and data generation for training. In contrast, our approach leverages complete environmental knowledge through Digital Twins, enabling direct gradient-based optimization without the computational overhead of exploration and learning phases required by RL methods.} Some studies incorporate obstacles for UAV route planning in sensor data collection \cite{9117104} and coverage optimization \cite{9614346}, but focus on single UAV operation or single target areas. While RL is useful when ABSs have access to limited environmental data, it demands extensive sensing and large-scale data generation for effective training. Furthermore, these works rely on stochastic channel models rather than precise RF propagation modeling. Recently, \cite{10486853} proposed a placement approach based on radio propagation maps and discretized 3D locations, but it doesn't consider navigation or interference caused by multiple UAVs deployment.

DTNs have gained interest as high-fidelity replicas of real-world networks scenarios \cite{9696282, 10121572, 10148936}, facilitating testing of UAV placements and communication technologies. Some studies explore DT-supported UAV resource allocation \cite{10870876} and network reconstruction \cite{9263396}, but comprehensive, license-free DTNs integrating accurate wireless propagation, client mobility, and system-level RAN control remain underdeveloped. {\reviewI  Table \ref{tab:feature_comparison} summarizes the features of proposed approach compared to related works presented in this section.}

\begin{table*}[t]
\centering
\begin{tabular}{lccccc}
\hline
\textbf{Autonomous ABS Feature / Capability} &
\makecell{\textbf{RL-based ABS}\\\textbf{deployment}\\\textbf{(e.g.\ \cite{10056676,9321160})}} &
\makecell{\textbf{Obstacle-aware UAV}\\\textbf{route / coverage}\\\textbf{(e.g.\ \cite{9117104,9614346})}} &
\makecell{\textbf{Radio-map-based}\\\textbf{placement}\\\textbf{\cite{10486853}}} &
\makecell{\textbf{DT-supported UAV}\\\textbf{deployment}\\\textbf{(e.g.\ \cite{10870876,9263396})}} &
\makecell{\textbf{Proposed} \\ \textbf{approach}} \\
\hline
Learning/optimization based & \cmark & \cmark & \cmark & \cmark & \cmark \\
Continuous-space UAV locations & \xmark & \cmark & \xmark & \xmark\textsuperscript{\cite{10870876}} \cmark\textsuperscript{\cite{9263396}}& \cmark \\
Handles physical obstacles & \xmark & \cmark & \cmark & \xmark & \cmark \\
Multi-UAV & \xmark & \xmark\textsuperscript{\cite{9117104}} \cmark\textsuperscript{\cite{9614346}} & \cmark & \cmark & \cmark \\
Multi-target-area coordination & \xmark & \cmark\textsuperscript{\cite{9117104}} \xmark\textsuperscript{\cite{9614346}}  & \cmark & \cmark & \cmark \\
Ray Tracing RF propagation modeling & \xmark & \xmark & \cmark & \xmark & \cmark \\
Navigation / dynamic repositioning & \xmark\textsuperscript{\cite{10056676}} \cmark\textsuperscript{\cite{9321160}} & \cmark & \xmark & \xmark\textsuperscript{\cite{10870876}} \cmark\textsuperscript{\cite{9263396}}  & \cmark \\
Interference management among multiple UAVs & \xmark & \xmark & \xmark & \cmark & \cmark \\
High-fidelity digital-twin network (DTN) & \xmark & \xmark & \xmark & \cmark & \cmark \\
Integrates clients mobility & \xmark & \xmark & \xmark & \xmark\textsuperscript{\cite{10870876}} \cmark\textsuperscript{\cite{9263396}} & \cmark \\
Allows system-level RAN simulation & \xmark & \xmark & \xmark & \xmark & \cmark \\
Free to use / open-source framework & \xmark & \xmark & \cmark & \xmark & \cmark \\
\hline
\end{tabular}

\caption{
{\reviewII  Feature comparison of existing ABS-related approaches versus ours. Proposed Multi-DT approach provides unique capabilities including continuous-space optimization, comprehensive obstacle handling, and differentiable RF propagation modeling that enable superior performance compared to learning-based alternatives that lack complete environmental information. Note that \cite{9263396} outlines a DT-based coordination framework, but does not propose any explicit autonomous UAV deployment solution.}}
\label{tab:feature_comparison}
\end{table*}


\begin{figure*}[t]
\centering
\begin{subfigure}{.5\textwidth}
  \centering
  \includegraphics[width=0.8\linewidth]{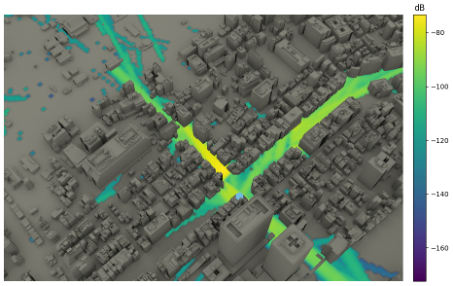}
  \caption{}
  \label{fig:sionna_tokyo}
\end{subfigure}%
\begin{subfigure}{.5\textwidth}
  \centering
  \includegraphics[width=0.87\linewidth]{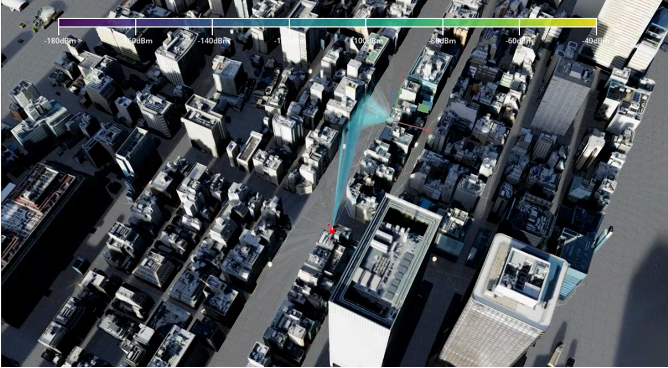}
  \caption{}
  \label{fig:aodt_tokyo}
\end{subfigure}%
\caption{The same Tokyo 3D map from high-detail PLATEAU dataset loaded in Sionna and AODT, used to demonstrate the proposed approaches for Multi-DT framework. (a) shows a path gain Coverage Map computed with Sionna and (b) presents a simulation frame from AODT multi-UE simulation in the same map location.}
\end{figure*}

\section{Bridging Sionna and AODT: A Unified Digital Twin Framework} 
\label{sec:multiDT}
The concept of using multiple DTs concurrently for a shared objective is still emerging \cite{CIMINO2021103501}. This work focuses on DTNs with Ray Tracing \cite{7152831} rather than statistical channel modeling, as they allow realistic multi-path propagation simulation in detailed 3D urban environments. We utilize two NVIDIA's DTNs: Sionna\footnote{Version 0.19, October 2024} \cite{10465179} and Aerial Omniverse Digital Twin (AODT)\footnote{Version 1.1.1, October 2024} \cite{AODT}. {\reviewI  Sionna and AODT represent a new class of AI-native, high-fidelity wireless simulation tools that go beyond the capabilities of traditional network simulators like NS-3~\cite{ns3} and OMNeT++~\cite{omnetpp}, or EM tools like Remcom Wireless InSite~\cite{remcom_wireless_insite} and ANSYS HFSS~\cite{ansys_hfss}. While traditional simulators focus on protocol-level abstraction or detailed electromagnetic modeling in static environments, Sionna and AODT integrate differentiable physical-layer models, photorealistic 3D environments, and efficient Ray Tracing simulation to support AI-driven design and optimization of 5G and 6G networks. This makes them uniquely suited for creating dynamic, end-to-end digital twins of urban wireless systems—enabling realistic channel modeling, beamforming, and users' mobility-aware optimization within a fully interactive environment. For researchers and engineers developing next-generation wireless technologies with AI at the core, Sionna and AODT offer a future-ready platform that bridges the gap between high-level network design, and low-level physical realities in a scalable, GPU-accelerated workflow.} This section summarizes their key features and highlights their design differences.

\noindent$\bullet$ \textbf{Sionna Ray Tracing (RT)}:
Sionna RT is part of NVIDIA's Sionna \cite{hoydis2022sionna} link-level simulation library. Its key feature is differentiability in RF simulation blocks, including statistical models and Ray Tracing, enabling direct optimization of network parameters and antenna orientation based on EM propagation effects. It leverages TensorFlow \cite{abadi2016tensorflow} for automatic differentiation and scalable gradient-based optimization. The Ray Tracing module utilizes the Mitsuba3 differentiable renderer \cite{mitsuba3}, built on Dr.Jit \cite{10.1145/3528223.3530099}, for efficient gradient computation. Fig. \ref{fig:sionna_tokyo} illustrates a Coverage Map generated with Sionna.

\noindent$\bullet$ \textbf{Aerial Omniverse Digital Twin (AODT)}: AODT, part of NVIDIA’s Omniverse DT ecosystem, supports EM propagation and system-level simulations. It enables realistic network deployment, leveraging NVIDIA Aerial CUDA Accelerated RAN \cite{NVIDIA_accel_RAN} for full GPU acceleration of 5G L1/L2 layers. Its high-performance Ray Tracing engine, written in C++/CUDA, outperforms Sionna’s Python-based implementation while maintaining functionally identical EM propagation effects. Though non-differentiable, AODT supports rapid multi-{\reviewI User Equipment (UE)} simulation data generation for offline analysis and ML/DL model training. This work focuses on AODT’s L1 EM simulation, leaving L2 integration for future studies. Fig. \ref{fig:aodt_tokyo} shows an EM simulation in AODT.

\noindent$\bullet$ \textbf{DTNs Choice Motivation}: 
Our motivation for combining these {\reviewI  specific} DTNs is as follows: (i) Sionna outperforms commercial competitors \cite{zhu2024toward}, while AODT supports multi-UE mobility, diverse antenna configurations, and efficient Ray Tracing; (ii) As shown in Table~\ref{tab:aodt_vs_sionna_comparison}, they offer distinct functionalities, such as differentiable tensor blocks in Sionna and system-level simulation in AODT; (iii) Both are freely available. 
While Sionna provides a differentiable Ray Tracer for gradient-descent optimization in multi-path propagation models, AODT supports large-scale simulations with higher Ray Tracing sampling and mobility features. As telecom operators explore next-generation networks, these DTNs offer complementary capabilities for innovative solutions. Integrating their outputs enables tackling complex challenges.

\begin{table}[t]
\centering
\begin{tabular}{|l|c|c|}
\hline
\textbf{Feature} & \textbf{AODT 1.1.1} & \textbf{Sionna 0.19} \\ \hline
Simulation target & System-level & Link-level \\ \hline
3D geometry format & OpenUSD & Mitsuba \\ \hline
PHY (L1) simulation & \checkmark & \checkmark \\ \hline
MAC (L2) simulation & \checkmark & - \\ \hline
5G waveform compliant & \checkmark & \checkmark \\ \hline
Multi-device (BS/UE) simulation & \checkmark & \checkmark \\ \hline
UE mobility engine & \checkmark & - \\ \hline
Coverage Maps & - & \checkmark \\ \hline
Differentiable & - & \checkmark \\ \hline \hline
\textbf{Ray Tracing simulation engine} &  &  \\ \hline
- Reflection & \checkmark & \checkmark \\ \hline
- Scattering / Diffusion & \checkmark & \checkmark \\ \hline
- Diffraction & \checkmark & \checkmark \\ \hline
- Surface material properties (ITU) & \checkmark & \checkmark \\ \hline
- Customizable antenna panels & \checkmark & \checkmark \\ \hline \hline
\textbf{Simulation Param. (max value)}  &  &  \\ \hline 
Num. of rays emitted at every RU & $1,000,000$ & \dag \\ \hline
Num. of reflection/diffusion events & $5$ & \dag \\ \hline
Num. of diffraction events & $1 \times~\text{path}$* & $1 \times~\text{path}$** \\ \hline
Num. of UE & $10,000$ & \dag \\ \hline
Num.of antenna elements per RU & $64$ & \S \\ \hline
Num. of antenna elements per UE & $8$ & \S \\ \hline

\end{tabular}
\caption{Simulation features and EM propagation effects capabilities of Aerial Omniverse Digital Twin (AODT) and Sionna. \dag = set by user (w/ GPU memory constraints);  \S = set by user (shared by all RUs/UEs in the simulation); * = for any interaction of the ray along its path; ** = only for LoS wedge interaction with transmitter. Note: Sionna quantities specifically relate to its Coverage Map function.}
\label{tab:aodt_vs_sionna_comparison}
\end{table}

\section{Challenges in Combining Digital Twins}
\label{sec:DTchallenges}

Integrating multiple DTNs presents several challenges due to their lack of built-in interoperability. While this work focuses on Sionna and AODT, these challenges apply broadly to other DTN combinations:

\begin{itemize}
    \item \textbf{Sharing 3D Urban Models}: Despite having similar Ray Tracing capabilities, Sionna and AODT use different scene descriptors (Mitsuba3 vs. OpenUSD). Although both support OpenStreetMap imports, high-resolution custom models require manual conversion.

    \item \textbf{Wireless Device Deployment Exchange}: AODT lacks procedural import/export functions for radio and user equipment placement, relying on manual GUI configuration. To bridge this gap, modifications were made to enable JSON-based deployment imports from Sionna and automate UE placement for specific areas in AODT.

    \item \textbf{Ray Tracing Variability}: Differences in stochastic Ray Tracing implementations, unit systems, and antenna models make direct comparison difficult. Parameter adjustments are necessary to align simulation outputs.

    \item \textbf{Different Simulation Features}: AODT supports only fixed ground stations, limiting ABS deployment studies. To overcome this, a custom BS mobility system was implemented, allowing pre-computed ABS trajectories during simulations.
\end{itemize}

Such challenges highlight the need for tailored solutions when integrating multiple DTNs for wireless network simulations and the main roadblocks addressed by our Multi-DT implementation.
{\reviewI 
\subsection{Shared Data Layer Components and Functions}
In order to address these challenges, we have implemented a dedicated Shared Data Layer that allows exchange of 3D models and deployment configurations and to easily validate complex wireless deployments via proposed Multi-DT platform. The Shared Data Layer architecture provides several critical advantages, which are rooted in the following design principles:
\begin{enumerate}
    \item \textbf{Platform Independence:} Each DT operates in its native environment while sharing standardized data representations;
    \item \textbf{Standardized Interoperability:} JSON-based protocols can accommodate different numbers of ABSs, AOIs, and UE configurations;
    \item \textbf{Multi-DT Deployment Adaptation:} Bidirectional communication enables dynamic scenario updates based on performance feedback from each DT;
    \item \textbf{Validation Integrity:} Cross-platform consistency checks ensure optimization results translate effectively between environments;
    \item \textbf{Extensibility:} Modular design allows integration of additional DT platforms and applications with minimal architectural changes.
\end{enumerate}

The Shared Data Layer implements a multi-faceted approach to handle the fundamental incompatibilities between Sionna and AODT platforms. It consists of four main functional components:
\begin{enumerate}
    \item \textbf{3D Scene Data Harmonization}: Sionna uses Mitsuba3 scene descriptors while AODT employs OpenUSD format, creating incompatibility for high-resolution 3D urban models. This component includes:
    \begin{itemize}
        \item \textbf{Custom Blender Script}: Automated conversion pipeline that reads OpenUSD scenes from AODT and exports Mitsuba3-compatible formats for Sionna;
        \item \textbf{Geometry Preservation}: Ensures building coordinates, surface materials properties, and structural details remain consistent across both platforms;
        \item \textbf{Coordinate System Alignment}: Maintains spatial consistency for Tokyo PLATEAU dataset across both environments by addressing possible unit measure differences (e.g., metric vs. imperial system) and coordinate system conventions (e.g. Global vs. Local, Right-handed vs. Left-handed Coordinate Systems).
    \end{itemize}

    \item \textbf{Device Deployment Configuration Exchange}: AODT lacks procedural import/export functions for radio equipment placement, relying on manual GUI configuration. Standardized data structures have been defined to exchange the following information via a JSON-based protocol:
    \begin{itemize}
        \item \textbf{AOIs Parameters}: Center coordinates $(z_k, w_k)$ and radius $r_k$ for each $k$-th AOI;
        \item \textbf{ABSs Parameters}: Coordinates $(x_i, y_i)$, orientations $(\phi_i, \theta_i)$, and transmission powers $P_i^{tx}$ for each $i$-th ABS;
        \item \textbf{ABSs Trajectories}: A list of $(x_i, y_i)$ coordinates for each $i$-th ABS at a given simulation time-step;
        \item \textbf{UEs Trajectories}: A list of coordinates $(x_u, y_u)$ for each $u$-th UE simulated in experiments with mobile terminals.
    \end{itemize}
    

    \item \textbf{Simulation Parameter Synchronization}: Different ray tracing implementations, unit systems, and antenna models make direct comparison difficult. Hence, several parameters need to be consistently tracked across Multi-DT platforms:
    \begin{itemize}
        \item \textbf{RF Parameter Alignment}: Center frequency ($f_c = 3.5$ GHz), sampling frequency and antenna patterns (TR 38.901 for ABS, half-wave dipole for UE);
        \item \textbf{Ray Tracing Harmonization}: Consistent material properties, types of ray interactions (e.g., specular reflections, diffusion, diffraction) and number of interactions limit per Ray;
        \item \textbf{Power Scale Matching}: Transmission power normalization (43.0 dBm baseline) and SIR computation standardization;
    \end{itemize}


    
    
\end{enumerate}

Fig. \ref{fig:multiDT_implementation} depicts the flow of data exchanged between the chosen DTs via implemented Shared Data Layer. {\reviewIII Table \ref{tab:all_parameters} provides a comprehensive list of all simulation parameters and notation used throughout this work.}
}
\begin{table}[t]
\centering
\resizebox{\columnwidth}{!}{%
\begin{tabular}{|c|l|c|c|}
\hline
\textbf{Parameter} & \textbf{Description} & \textbf{\makecell{Equation \\ / Section}} & \textbf{Value} \\
\hline
\multicolumn{4}{|c|}{\textbf{Loss Function Components}} \\
\hline
$L_p$ & Total loss function to minimize & (1) & - \\
$\alpha$ & Scaling factor for coverage term & (1) & 0.01 \\
$\beta$ & Scaling factor for attraction penalty & (1) & 1.0 \\
$\gamma$ & Scaling factor for repulsion penalty & (1) & 0.8 \\
$\eta$ & Scaling factor for collision penalty & (1) & 1.0 \\
$K$ & Coverage factor for spatial distribution & (2) & - \\
$P_u$ & Repulsion penalty between ABSs & (3) & - \\
$P_a$ & Attraction penalty toward AOIs & (4) & - \\
$P_b$ & Collision avoidance penalty & (6) & - \\
\hline
\multicolumn{4}{|c|}{\textbf{Spatial and Geometric Parameters}} \\
\hline
$N$ & Total number of ABSs & (2,3) & 10 \\
$M$ & Total number of Areas of Interest (AOIs) & (4) & 5 \\
$p_i$ & Position coordinates $(x_i, y_i)$ of $i$-th ABS & (2,3) & - \\
$c_k$ & Center coordinates $(z_k, w_k)$ of $k$-th AOI & (4) & {\reviewIII Table \ref{tab:AOIs_config}} \\
$r_k$ & Radius of $k$-th AOI & (4) & 250-300m \\
$G$ & Set of evenly spaced 2D grid coordinates & (2) & $5 \times 5$ grid \\
$g$ & Individual grid point coordinate & (2) & - \\
$m_e$ & Margin distance from map edges & (2) & 150m \\
$d_{min}$ & Minimum distance between ABSs & (3) & 400m \\
\hline
\multicolumn{4}{|c|}{\textbf{Building and Collision Parameters}} \\
\hline
$B$ & Number of buildings in the map & (6) & * \\
$h$ & ABSs' elevation & V-A & 70m \\
$c_b$ & Minimum allowed distance from buildings & (6) & 15m \\
$d_{ib}$ & Distance from ABS $i$ to building $b$ & (6,7) & - \\
$(p_x, p_y)$ & ABS XY coordinates & (7) & - \\
$(m_x, m_y)$ & Min. XY coords. of building bounding box & (7) & - \\
$(M_x, M_y)$ & Max. XY coords. of building bounding box & (7) & - \\
$d_x, d_y$ & X,Y components of distance to building & (7) & - \\
\hline
\multicolumn{4}{|c|}{\textbf{Steepness and Weight Parameters}} \\
\hline
$\kappa_a$ & Steepness factor for attraction exponential & (4) & 0.02 \\
$\kappa_b$ & Steepness factor for collision penalty & (6) & 0.5 \\
$\kappa_i$ & Steepness factor for sigmoid function & (5) & 0.25 \\
$\omega_k$ & Weight factor for $k$-th AOI & (4,5) & - \\
$\sigma(z,t,\kappa)$ & Modified sigmoid: $\frac{1}{1+e^{-\kappa(z-t)}}$ & (5) & - \\
\hline
\multicolumn{4}{|c|}{\textbf{Coverage Map and Ray Tracing}} \\
\hline
$C$ & Coverage map tensor $\mathbb{R}^{N \times C_x \times C_y}$ & (8) & - \\ 
$C_{i,x,y}$ & Coverage map cell for ABS $i$ at $(x,y)$ & (8) & - \\
$\hat{g}(i,x,y)$ & Estimated path gain for ABS $i$ at cell $(x,y)$ & (8) & - \\
$N_R$ & Number of valid rays hitting cell $(x,y)$ & (8) & * \\ 
$N_C$ & Total coverage map cells: $C_x \cdot C_y$ & (8) & * \\
$|h(s(\psi_n))|^2$ & Squared amplitude of path coefficients & (8) & - \\
$r(\psi_n)$ & Length of $n$-th path with direction $\psi_n$ & (8) & - \\
$\alpha(\psi_n)$ & Angle between map normal and arrival direction & (8) & - \\
$s(\psi_n)$ & Intersection point of $n$-th path with map & (8) & - \\
\hline
\multicolumn{4}{|c|}{\textbf{Signal Strength and Interference}} \\
\hline
$P_i^{tx}$ & Initial transmission power of $i$-th ABS (dBm) & (8,9) & 43.0 dBm \\
$RSS_{i,x,y}$ & Received Signal Strength at cell $(x,y)$ & - & - \\
$R_{i,x,y}$ & Signal-to-Interference Ratio at cell $(x,y)$ & (9) & - \\
$\epsilon$ & Noise floor (fixed) & (9) & $1 \times 10^{-20}$ \\
$M_{i,x,y}$ & Coverage mask for ABS $i$ at cell $(x,y)$ & (10) & 0 or 1 \\
$\hat{r}_i$ & Effective average SIR for $i$-th ABS (dB) & (11) & - \\
\hline
\multicolumn{4}{|c|}{\textbf{Optimization Loss Functions}} \\
\hline
$\mathcal{L}_{smoothmin}$ & Smooth minimum approximation loss & (12) & - \\
$\mathcal{L}_{avgSIR}$ & Average SIR maximization loss & (14) & - \\
$\mathcal{L}_o$ & Total orientation optimization loss & (15) & - \\
$\mathcal{L}_u$ & Unweighted AOI SIR optimization loss & (18) & - \\
$\mathcal{L}_w$ & Weighted AOI SIR optimization loss & (19) & - \\
$NLSE(r, \beta_L)$ & Negative Log-Sum-Exp function & (13) & - \\
$LSE(r_m^*, \beta_L)$ & Log-Sum-Exp for $m$-th AOI & (17) & - \\
\hline
\multicolumn{4}{|c|}{\textbf{Algorithm and Optimization Parameters}} \\
\hline
$\beta_L$ & Temperature for NLSE/LSE functions & (12,13) & 1.0 \\
$\xi$ & Scaling factor for average SIR term & (15) & 0.25 \\
$w_m$ & Softmin weight for $m$-th AOI & (20) & - \\
$T$ & Temperature parameter for softmin & (20) & 25 \\
$\phi_i$ & Mechanical azimuth of $i$-th ABS (degrees) & V-B & $\left[-2\pi, 2\pi\right]$ \\
$\theta_i$ & Mechanical tilt of $i$-th ABS (degrees) & V-B & $\left[\pi/7, 6\pi/7\right]$ \\
\hline
\end{tabular}%
}
\caption{{\reviewI  Complete Parameter Reference for ABS Deployment Optimization Framework. * = Map-dependent.}}
\label{tab:all_parameters}
\end{table}
\section{Gradient-Based ABSs Deployment Optimization}
\label{sec:ABSoptimization}

This section presents a case study using the proposed Multi-DT framework to develop an autonomous ABS deployment algorithm for enhancing network capacity or restoring coverage in disaster-affected areas. The key requirements for this algorithm are as follows:  

\begin{itemize}
    \item ABSs need to be deployed rapidly and autonomously;
    \item These deployments are temporary and adaptive, with possibility of quick change of objectives and targets;
    \item The deployment algorithm has to be general enough to adapt to different urban scenarios and be aware of obstacles in 3D space during navigation;
    \item The deployment algorithm has to be aware of wireless propagation properties of the environment to avoid incurring in undesirable interference once hovering locations are identified.
\end{itemize}

By exploiting the rich information available within the DTs, we employ gradient-based optimization to first generate ABS navigation routes from random initial positions, targeting multiple coverage areas, and then optimize  orientation and transmission power to minimize interference, using Sionna's differentiable Ray Tracer. This approach enables a flexible method that can easily be adapted to different urban scenarios and coverage requirements, without the costly data generation and training required for learning-based methods. To validate our approach, we use Tokyo’s high-resolution 3D map from PLATEAU \cite{PLATEAU} dataset provided by AODT. The geometry is converted from OpenUSD to Mitsuba3 format for compatibility with Sionna, where the optimization is implemented via TensorFlow.

\subsection{Location Optimization}
\label{sec:ABSlocation}

We assume ABSs hover at a fixed elevation of {\reviewI $h = 70$m,  which is considered a hyper-parameter, and we focus on optimizing ABSs' locations only} in the XY-plane. {\reviewI  This elevation was selected to ensure ABSs operate above most urban obstacles in the targeted map while maintaining practical deployment constraints.} The deployment optimization aims to find optimal locations $(x_i,y_i)$ for $N$ ABSs to cover $M$ Areas of Interest (AOIs), each defined by $(z_m,w_m)$ and radius $r_m$. Initially, ABSs are placed semi-randomly\footnote{For initial deployment, we randomly samples initial positions for each ABSs while ensuring they are not generated within buildings/obstacles and do not cluster too closely.}, as they may be deployed from aerial vehicles, storage hubs, or accessible zones near disaster-affected areas. Optimal paths are computed to navigate from initial positions to target locations while avoiding obstacles and collisions. No limit is imposed on the number of ABSs per AOI, allowing flexibility in coverage. Once AOIs are served, remaining ABSs redistribute in their surroundings to enhance coverage or act as relays. Simultaneously computing optimal routes and locations requires tight coordination among ABSs. To achieve this, we model location optimization using a Particle Swarm Optimization-inspired gradient-descent approach \cite{NOEL2012353}. By embedding ABS interactions with the environment into the loss function, the algorithm directly samples navigation waypoints, guiding ABSs through the optimization landscape.

Specifically, we define our loss function to be minimized as a composition of multiple loss terms summed together, with each individual loss addressing one of our optimization criteria. The devised ABS deployment optimization function is determined as follows:
\begin{equation}
    \mathcal{L}_p = -\alpha K + \beta P_a + \gamma P_u + \eta P_b
\end{equation}

where $K$ corresponds to a coverage factor, $P_u$ is the repulsion penalty, $P_a$ is the attraction penalty and $P_b$ is a collision avoidance penalty. $\alpha$, $\beta$, $\gamma$ and $\eta$ correspond to scaling factors for each of the loss terms and are considered as hyper-parameters. 

$K$ and $P_u$ together encourage even distribution of particles across maps. Specifically, $K$ is defined as follows:
\begin{equation}
    K=\sum_{\boldsymbol{g} \in \boldsymbol{G}} \min _{1 \leq i \leq N}\left\|\boldsymbol{g}-\boldsymbol{p}_i\right\|
\end{equation}
 and $P_u$ is defined as:
\begin{equation}
    \small{P_u=\sum_j^N \sum_i^N \max \left(0, d_{\min }-\left\|\boldsymbol{p}_i-\boldsymbol{p}_j\right\|\right)}
\end{equation}
where $\boldsymbol{G}$ is a set of $\boldsymbol{g}$ evenly spaced 2D grid coordinates across the XY-plane of the 3D map and $\boldsymbol{p}_i = (x_i, y_i)$ is the coordinates of the $i$-th ABS. The coordinates and size of the grid points is computed by taking in input the number of reference points along the X and Y-axis and evenly distributing them across the map, leaving a margin of $m_e$ meters from the edges of the ground plane. These two terms together aims to maximize the sum of minimum distances from each grid points to all ABSs location while keeping the distance among each ABS at a minimum of $d_{\min}$, which is also considered a hyper-parameter.

The attraction penalty $P_a$ aims to pull ABSs in the map towards AOIs and is defined as follows:
\begin{equation}
\begin{split}
     P_a=\sum_{k=1}^M \sum_{i=1}^N \left[ \omega_k \cdot\left\|\boldsymbol{p}_i-\boldsymbol{c}_k\right\|-\left(1-\omega_k\right) \cdot \right. \\
     \left. \exp \left(-\kappa_a \cdot\left\|\boldsymbol{p}_i-\boldsymbol{c}_k\right\|-r_k\right)\right]
\end{split}
\end{equation}

where $\boldsymbol{c}_k$ and $r_k$ corresponds to the center coordinates and radius values of the $k$-th AOI, $\kappa_a$ is a hyper-parameter steepness factor for the exponential function and 
\begin{equation}
    \small{\omega_k=\exp \left[-\sum_{i=1}^N \sigma\left(\left\|\boldsymbol{p}_i-\boldsymbol{c}_k\right\|, \frac{2 r_k}{3}, -\kappa_i\right)\right]}
\end{equation}
 is a weight factor that decreases as more ABSs end up within a given AOI. To avoid harsh discontinuity in the loss function, we use $\sigma(z, t, \kappa) = \frac{1}{1 + e^{-\kappa\left(z - t\right)}}$ as a modified sigmoid function to smoothly approximate the condition $\left\|\boldsymbol{p}_i-\boldsymbol{c}_k\right\| < \frac{2 r_k}{3}$ and $\kappa_i$ is a steepness factor. The idea is that, for each AOI, all ABSs are linearly attracted by it and as soon as one or more ABS are within $2/3$ of the AOI radius, the attraction switches to exponential to increase the pull of those ABSs toward the center of the AOI, while all the other particles pulls will be “switched off” for that particular AOI. 

\begin{figure}[t]
\centering
\includegraphics[width=0.9\columnwidth]{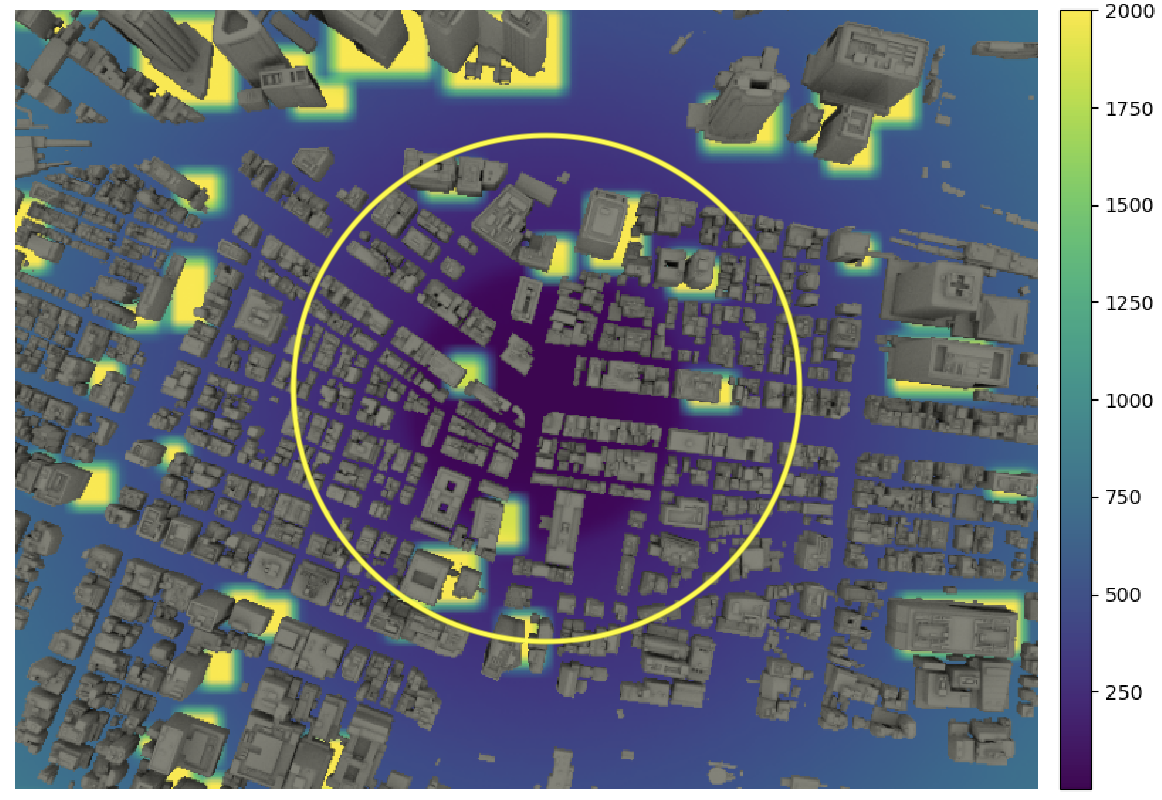}
\caption{Visualization of loss function terms $\gamma P_a + \eta P_b$ projected over the 2D ground-plane of a 3D map of Tokyo for a sample AOI (delimited by the yellow circle) {\reviewI  and considering ABSs' elevation of $h = 70$m}. In this visualization, only one ABS is considered and loss values are clipped in the range $[0,2000]$. }
\label{fig:loss_visual}
\end{figure}

Finally, the collision avoidance penalty $P_b$ is defined as follows:
\begin{equation}
    P_b=\sum_i^{N} \sum_b^B \exp \left(\kappa_b \cdot\left(-d_{ib}+c_b\right)\right)
\end{equation}
where $B$ is the number of buildings in the map, $\kappa_b$ is a steepness factor, $c_b$ is the minimum allowed distance in meters from a building, $d_{ib}$ is the distance function from any given ABS $i$ to any given building $b$'s closest edge, defined as 
\begin{equation}
\begin{matrix}
d_{ib} = \sqrt{d_x^2 + d_y^2}, \\
d_x = \max\left(\max\left(m_x - p_x, p_x - M_x\right), 0 \right), \\
d_y = \max\left(\max\left(m_y - p_y, p_y - M_y\right), 0 \right)
\end{matrix}
\end{equation}
given $(p_x,p_y)$ as the ABS's XY coordinates, $(m_x, m_y)$ and $(M_x, M_y)$ as the minimum and maximum XY coordinate of a building's binding box, respectively. As the ABS elevation is assumed constant, buildings that have heights below the hovering elevation (plus a tolerance of $15$m to avoid blockages from rooftops) are excluded from the penalty term computation. In order to provide a better understanding of how the proposed loss function works, Fig. \ref{fig:loss_visual} offers a visual representation of the environment-dependent loss terms $\gamma P_a + \eta P_b$. From this figure, it is possible to note how the loss values progressively become smaller for locations closer to the center of AOI, while higher values are present in proximity of buildings that have height higher than defined ABS' hovering height.

The optimization procedure considers the $(x_i, y_i)$ coordinates for all the $N$ ABSs as a set of parameters $\Theta_{\text{l}}$, and aims to perform gradient-descent optimization by following the inverse direction provided by the gradients of the loss function w.r.t. the location parameters, i.e.  $\frac{\delta \mathcal{L}_p}{\delta \Theta_{\text{l}}}$.

\subsection{Orientation and Power Optimization}
\label{sec:ABS_orient_pow_optim}
Once the routes and final locations of ABS have been identified by the previous optimization step, their orientations also need to be adjusted. For this optimization step, we initially assume that the ABS antenna panels face directly the ground plane, with a mechanical tilt equal to $\theta_i = 90^{\circ}$ for each $i$-th ABS. As their location in the 3D space will also affect the wireless propagation features of the serviced area, their deployment needs to be optimized so that their transmissions will not interfere with each other, hence by minimizing their mutual interference by jointly adjusting their location and transmission power.

To do so, we utilize the differentiable Ray Tacer provided by Sionna to compute a differentiable Coverage Map $\boldsymbol{C} \in \mathbb{R}^{N \times C_x \times C_y}$, which consists of the average path gain experienced at each $(x,y)$ location of a discretized version of the ground plane made of $N_C = C_x \cdot C_y$ surface cells of equal size. Each cell reports a value equal to the sum of contributions of reflected/diffused paths and diffracted paths, for each of the $N$ ABS. For a given ABS $i$, the value of each cell $\boldsymbol{C}_{i,x,y} = \hat{g}(i, x, y)$ is computed via Monte Carlo simulation as follows:
\begin{equation}
    \small{\hat{g}(i, x, y)=\frac{4 \pi}{N_RN_C} \sum_{n=1}^{N_R}\left|h\left(s\left(\psi_n\right)\right)\right|^2 \frac{r\left(\psi_n\right)^2}{\left|\cos \alpha\left(\psi_n\right)\right|} \mathbf{1}{\left\{s\left(\psi_n\right) \in C_{i, x, y}\right\}}}
\end{equation}
where $N_R$ is the number of valid rays hitting  $\boldsymbol{C}_{i,x,y}$ cell,  $\left|h\left(s\left(\psi_n\right)\right)\right|^2$ is the squared amplitude of the path coefficients at position $s(\psi_n)$ (i.e., the point where the $n$-th path with direction of departure $\psi_n$ intersects the Coverage Map), $r(\psi_n)$ is the lenght of the $n$-th path with direction of departure $\psi_n$, $\alpha(\psi_n)$ is the angle between the coverage map normal and the direction of arrival of the path with direction of departure $\psi_n$, and $\mathbf{1}\left\{s\left(\psi_n\right) \in C_{i, x, y}\right\}$ is the function that sets a value of $1$ if the intersection point with the Coverage Map is within the current $(x,y)$ cell, or zero otherwise. If we consider a transmission power $P_i^{\text{tx}}$ expressed in Watts for each $i$-th ABS, we can obtain the Received Signal Strength (RSS) at each receiver location in the map as $\text{RSS}_{i,x,y} = P_i^{\text{tx}} \cdot \hat{g}(i, x, y)$.


We propose two separate optimization strategies based on gradient-descent in order to (i) improve the mutual interference of the ABSs over the whole map and (ii) improve the mutual interference over a set of specific AOIs. 
\subsubsection{ABSs' mutual interference optimization (Method 1)}
\label{sec:ABSorientation_overall}
From proposed location optimization approach, we observe that the ABSs tend to group around the AOIs in a clustered formation, assuming AOIs are sufficiently close to each other. Intuitively, the ABSs will have lower {\reviewI  Signal-to-Interference-Ratio (SIR)} when located within the cluster and higher SIR when located on its borders. Although it might seem reasonable to optimize orientations by maximizing the average SIR for all devices in the map, it is important to note that this approach might not necessarily produce a better configuration, as the gradients might favor devices with excessively high SIR while neglecting the ones in the lowest SIR regions, creating areas in the map with a wider disparity of Quality-of-Service (QoS). Hence, we formulate a loss function based on a Max-Min approach that prioritizes improving the SIR of ABSs that experience the highest interference (i.e., those within the ABS cluster) while making sure not to excessively disrupt those in higher SIR regions (i.e., the ones on the cluster border). By targeting the worst-case SIR, the Max-Min approach ensures fairness by uplifting the least-performing areas while avoiding over-optimization of ABSs with dominant SIR conditions. Specifically, for each $i$-th ABS we compute its SIR map in linear scale $\boldsymbol{R}_i \in \mathbb{R}^{C_x \times C_y}$ from the RSS perceived at each cell location by combining the coverage map $\boldsymbol{C} \in \mathbb{R}^{N \times C_x \times C_y}$ path gains produced by Sionna and the transmission power of each ABS as follows:
\begin{equation}
\boldsymbol{R}_{i,x,y} = \frac{P_i^{\text{tx}} \boldsymbol{C}_{i,x,y}}{\sum_{j \neq i} P_j^{\text{tx}} \boldsymbol{C}_{j,x,y} + \epsilon}
\end{equation}
where $\boldsymbol{C}_{i} \in \mathbb{R}^{\boldsymbol{C}_x \times \boldsymbol{C}_y}$ is the $i$-th ABS's coverage map and $\epsilon = 1e-20$ is a small value used as a proxy for thermal noise term and for numerical stability. 
Then, we compute a coverage mask $\boldsymbol{M}_i \in \mathbb{R}^{C_x \times C_y}$:
\begin{equation}
\boldsymbol{M}_{i,x,y} = \begin{cases}
    1, & \text{if } \boldsymbol{C}_{i,x,y} > 0\\
    0,              & \text{otherwise}
    \end{cases}
\end{equation}
that we use to compute the average effective SIR $\hat{r}_i \in \mathbb{R}$ expressed in dB as follows:
\begin{equation}
    \hat{r}_i = \frac{\sum_y\sum_x \boldsymbol{R}_{i,x,y}^{dB} \cdot \boldsymbol{M}_{i,x,y}}{\sum_y\sum_x \boldsymbol{M}_{i,x,y}}
\label{eq:effective_sir}
\end{equation}
where $\boldsymbol{R}_i^{dB} = 10 \cdot \text{log}_{10}\left(\boldsymbol{R}_i\right)$ correspond to the SIR map $\boldsymbol{R}_i$ in logarithmic scale. It is important to note that the effective SIR, rather than the SIR over the whole map, has to be computed in order to avoid diluting the average SIR computation over the cells that have no coverage.

Once all the $ \boldsymbol{r} = \left\{\hat{r}_1, ... , \hat{r}_N \right\}$ are obtained for all $N$ ABSs, we formulate the first term of our loss function as follows:
\begin{equation}
    \mathcal{L}_{\text{smoothmin}} = -\text{NLSE}\left(\boldsymbol{r}, \beta_{\text{L}} \right)
\end{equation}

where the Negative Log-Sum-Exp (NLSE) is used a smooth approximation of the \textit{minimum} function. NLSE is defined as:
\begin{equation}
    NLSE(\boldsymbol{r}, \beta_{\text{L}}) = -\frac{1}{\beta_{\text{L}}} \text{log}\left( \sum_{m=1}^M e^{-\beta_{\text{L}} r_m}\right)
\end{equation}
and for small values of $\beta_{\text{L}} > 0$ it progressively includes more values of the input vector in the computation of approximate minimum, avoiding steep discontinuities in the loss output as it maximizes its minimums. Moreover, in order to avoid degrading too much the SIR of the other ABS, we add a second term to our loss function defined as follows:
\begin{equation}
    \mathcal{L}_{\text{avgSIR}} = - \frac{\sum_m^M r_m}{M}
\end{equation}

 that aim to maximize the overall average SIR for all ABSs and that is intended to be scaled using a factor $0 < \xi < 1$ to avoid over-optimization of ABS with high SIR as explained before.

 Finally, we obtain the total loss function for orientation optimization by putting together the two loss terms defined above as follows:
 \vspace{-1em}
 \begin{equation}
     \mathcal{L}_o = \mathcal{L}_{\text{smoothmin}} + \xi \mathcal{L}_{\text{avgSIR}}
 \end{equation}
 The optimization procedure considers the set of $(\phi_i, \theta_i, P_i^{\text{tx}})$ mechanical azimuth, mechanical tilt and transmission power for all the $N$ ABSs as a set of parameters $\Theta_{\text{op}}$, and aims to perform gradient-descent optimization by following the inverse direction provided by the gradients of the loss function w.r.t. the location parameters, i.e.  $\frac{\delta \mathcal{L}_o}{\delta \Theta_{\text{op}}}$.

\subsubsection{AOIs' SIR optimization (Method 2)}
\label{sec:ABSorientation_AOI}
The second approach we propose focuses on improving the mutual interference of ABSs specifically for a set of AOIs considered in the targeted urban scenario: while the previous approach considers the entirety of covered cells in the map, this approach focuses on maximizing the effective SIR of cells associated with the AOIs identified by the network operator by \textit{maximizing the SIR of each AOI's serving ABS}, in order to improve the signal strength of the UEs located in those areas.

To formulate the loss function for this strategy, we refer to the effective average SIR for each $i$-th ABS from Eq. \ref{eq:effective_sir}. In this case, instead of considering the entire coverage map to compute $\hat{r}_i$, we only consider the square area described by the center $(z_m, w_m)$ of the $m$-th AOI and defined within $\pm r_m$ range of its radius on both X and Y-axis. To do so, we first obtain the cell's $x$ and $y$ indexes of AOI's center, $x_m^*$ and $y_m^*$. Then, we compute the length in cells associated with AOI's radius, $r_m^*$, assuming equal cell size along the X and Y-axis. Finally, we extract the coverage map's area of $m$-th AOI, $\boldsymbol{A}_i \in \mathbb{R}^{2 r_m^* \times 2 r_m^*}$ for each $i$-th ABS:
\begin{equation}
    \boldsymbol{A}_i = \boldsymbol{R}_{i,x_m^* \pm r_m^*, y_m^* \pm r_m^*}
\end{equation}

Once $\boldsymbol{A}_i$ is obtained, we compute the effective SIR for all ABS $\boldsymbol{r}_m^* = \{\tilde{r}_1, ..., \tilde{r}_N\}$ using Eq. \ref{eq:effective_sir} and by substituting $\boldsymbol{R}_i$ with $\boldsymbol{A}_i$. We then obtain its \textit{smooth maximum} using Log-Sum-Exp (LSE) function, defined as:
\begin{equation}
    LSE(\boldsymbol{r}^*_m, \beta_{\text{L}}) = -NLSE(\boldsymbol{r}^*_m, \beta_{\text{L}})
    \label{eq:max_AOI_SIR}
\end{equation}

The goal is to optimize the orientations and transmission powers of all ABSs in order to \textit{maximize} the effective SIR for the ABS serving a given AOI (i.e., the one that has highest SIR for the cells corresponding to a given AOI), for all $M$ AOIs. Combining these terms for all AOIs, we obtain the following unweighted loss term:
\begin{equation}
    \mathcal{L}_{u} = -\sum_{m=1}^M LSE(\boldsymbol{r}_m^*, \beta_{\text{L}})
    \label{eq:unweighted_AOI_loss}
\end{equation}

Similarly to the approach discussed in \ref{sec:ABSorientation_overall}, we want to prioritize optimization of AOIs that suffer from the highest SIR compared to all others, while still aiming to improve collectively the SIR experienced in all AOIs. To do so, we use a weighted version of Eq. \ref{eq:unweighted_AOI_loss} that uses \textit{softmin with temperature} function to assign priorities to each SIR maximization objective. Specifically, the weighted loss function will look as follows:
\begin{equation}
\mathcal{L}_{w} = -\sum_{m=1}^{M} w_m \cdot LSE(\boldsymbol{r}_m^*, \beta_{\text{L}})
\end{equation}
where the weights $w_m$ sum up to $1.0$ and are defined using the softmin function:
\begin{equation}
w_m = \frac{\exp\left(-\frac{x_m}{T}\right)}{\sum_{j=1}^{M} \exp\left(-\frac{x_j}{T}\right)}
\end{equation}
where $x_m = LSE(\boldsymbol{r}_m^*, \beta_{\text{L}})$ and $T$ is the temperature hyper-parameter to control the sharpness of the weight distribution:
a \textit{lower} temperature makes the softmin more sensitive to differences, giving much higher weights to smaller values, while a \textit{higher} temperature smooths the weights, distributing attention more evenly across all values.
In order to promote fairness while still prioritizing the AOI with lowest perceived SIR, we choose a high temperature temperature approach.

In this case, we aim to optimize the same set of parameters $\Theta_{\text{op}}$ introduced in the previous section, but this time optimizing w.r.t. $\mathcal{L}_w$ loss function, i.e. $\frac{\delta \mathcal{L}_w}{\delta \Theta_{\text{op}}}$.

\begin{table}[t]
\centering
\begin{tabular}{|l|l|l|l|l|l|}
\hline
\textbf{Parameter} & \textbf{AOI 0} & \textbf{AOI 1} & \textbf{AOI 2} & \textbf{AOI 3} & \textbf{AOI 4} \\ \hline
$z_m$  & 450.0  & -247.0 & -423.0 & 353.0  & -852.0 \\ \hline
$w_m$  & 168.0  & 145.0  & -416.0 & -622.0 & 133.0  \\ \hline
$r_m$  & 300.0  & 250.0  & 250.0  & 250.0  & 250.0  \\ \hline
\end{tabular}
\caption{Configuration of Areas of Interest (AOIs) for experimental evaluation of gradient-based ABS deployment. Coordinates $(z_m, w_m)$ and radius $r_m$ of the $m$-th AOI are expressed in meters.}
\label{tab:AOIs_config}
\end{table}
\vspace{-0.5em}
\subsection{Performance Evaluation}
{\reviewII 
\subsubsection{Hyperparameter Selection Methodology}

Our hyperparameter values were determined through systematic exploration guided by the following principles:
\begin{itemize}
    \item \textbf{Loss Function Scaling Factors:} Values selected to balance competing optimization objectives while ensuring numerical stability. $\alpha = 0.01$ provides lower weight for coverage to prevent overshadowing obstacle avoidance, while $\beta = 1.0$ ensures standard weight for AOI attraction. $\gamma = 0.8$ allows necessary clustering near AOIs, and $\eta = 1.0$ maintains full weight for safety-critical collision avoidance.
    \item \textbf{Steepness Factors:} $\kappa_a = 0.02$ provides smooth attraction gradients avoiding optimization instability, $\kappa_b = 0.5$ creates sharp building penalties while maintaining differentiability, and $\kappa_i = 0.25$ ensures smooth sigmoid transitions.
    \item \textbf{Optimization Parameters:} $\beta_L = 1.0$ provides appropriate smoothness for minimum approximation, $\xi = 0.25$ prevents over-optimization of high-SIR ABSs, and $T = 25$ ensures fair attention distribution across AOIs.
\end{itemize}

These values were validated by achieving over 97\% AOI satisfaction rates and effective obstacle avoidance across 1,800+ test runs.
}
\subsubsection{Results for ABS Positioning}
\label{sec:ABSlocation_perf}
In order to evaluate the performance of proposed gradient-descent based route finding and positioning algorithm described in Sec. \ref{sec:ABSlocation}, we define a fixed set of $M=5$ non-overlapping AOIs in the area of Tokyo 3D map described above. Table \ref{tab:AOIs_config} reports the configuration chosen for this experiments. 
We define an \textit{AOI satisfaction rate}, defined as:
\begin{equation}
S_{AOI} = \frac{\sum_{m=1}^M \min \left( 1, \sum_{n=1}^N  \mathbf{1}\left\{ d_n \leq \frac{2r_m}{3} \right\} \right)}{M} 
\end{equation}
which determines the ratio of correctly served AOIs by checking that distance $d_n = || \boldsymbol{p}_n - \boldsymbol{c}_m ||$ from the ABS coordinate $\boldsymbol{p}_n = (x_n, y_n)$ obtained at the end of the optimization to any given AOI center $\boldsymbol{c}_m = (z_m, w_n)$ coordinate is within $2/3$ of its radius $r_m$, while allowing for one or more ABS to hover within the same AOI. Moreover, in order to test the performance of proposed algorithm under different conditions, we evaluate this approach for different number of AOIs considered at once in the map. We perform multiple experiments with $m=\{1, 2, 3, 4, 5\}$ and for each of these configuration we test all the $\binom{M}{m}$ possible combinations of AOIs. Finally, for each AOI configuration, we perform $50$ tests with initial semi-random deployment of ABSs. For each experiment, we consider a number of ABS $N=10$ and use Adam optimizer with a learning rate $l_r=2.0$, a limit of $2500$ optimization iterations and early stopping criterion with a patience of $20$ training epochs. We use hyper-parameter exploration to define our loss parameters: we choose loss term scaling factors $\alpha = 0.01$, $\beta = 1.0$, $\gamma = 0.8$ and $\eta = 1.0$; for steepness factors, we select $\kappa_a = 0.02$, $\kappa_b = 0.5$, $\kappa_i = 0.25$; finally, we define $c_b = 15$m and $d_{\text{min}} = 400$m as minimum distances of ABSs from buildings and among ABSs themselves, respectively, and configure a set of $5\times5$ grid points equally distributed along the X and Y-axis of the map with a margin $m_e=150$m from its edges, used for the coverage term $K$.



\begin{figure}[t]
\centering
\includegraphics[width=0.95\columnwidth]{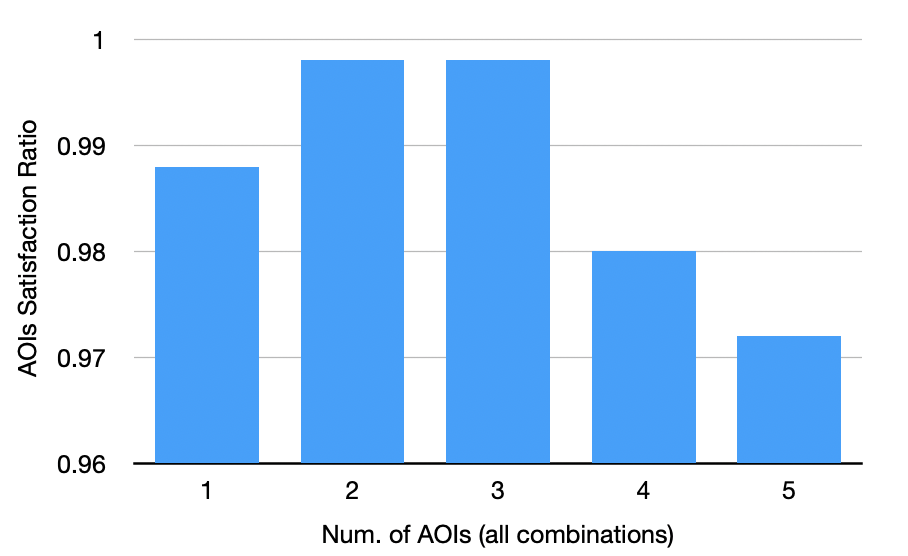}
\caption{AOI satisfaction metric for all combinations of $M$ AOIs taken in groups of $m=\{1, 2, 3, 4, 5\}$. Each combination is averaged over 50 runs with semi-random initial ABS deployment, for a total of $\{250, 500, 500, 250, 50\}$ runs each. }
\label{fig:AOI_satisfaction}
\end{figure}

Fig. \ref{fig:AOI_satisfaction} shows the average satisfaction rate for all AOI combinations. The results indicate that our approach successfully configures ABS deployments, achieving an AOI satisfaction rate of over $97\%$ across all runs and configurations. This demonstrates its effectiveness in navigating obstacles and landing in designated service areas.

\begin{figure*}[t]
\centering
\begin{subfigure}{.33\textwidth}
  \centering
  \includegraphics[width=1.15\columnwidth]{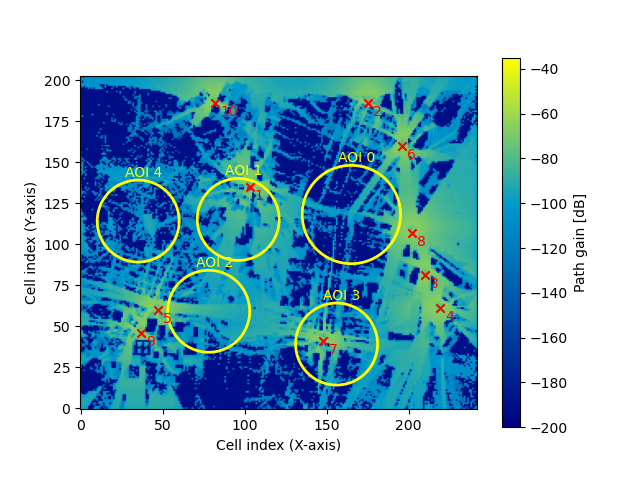}
  \caption{Initial random deployment}
\label{fig:ABS_init_deployment}
\end{subfigure}%
\hspace{-2.2em}
\begin{subfigure}{.33\textwidth}
  \centering
  \includegraphics[width=1.15\columnwidth]{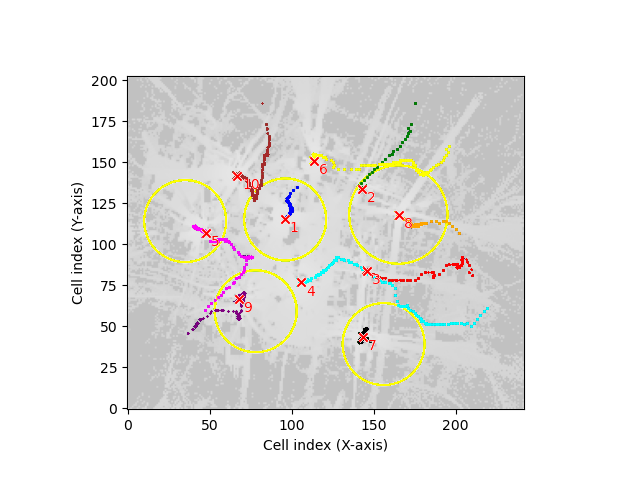}
  \caption{Generated routes}
  \label{fig:ABS_routes}
\end{subfigure}%
\begin{subfigure}{.33\textwidth}
  \centering
  \includegraphics[width=1.15\columnwidth]{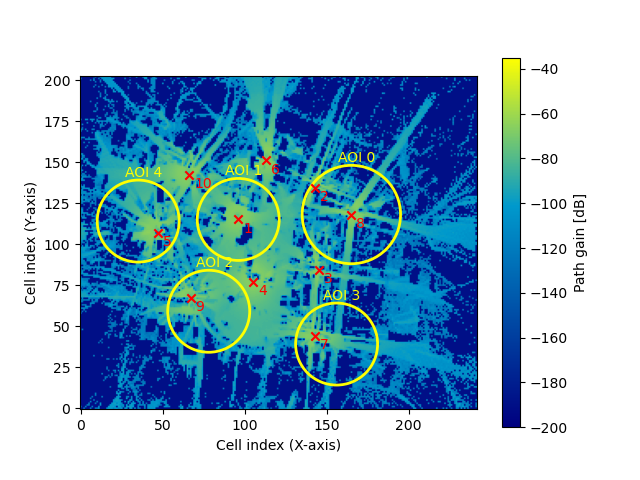}
  \caption{Final deployment}
  \label{fig:ABS_final_deployment}
\end{subfigure}%
\caption{Sample run from gradient-based ABS deployment optimization with $M=5$ AOIs and $N=10$ ABSs. Each ABS trajectory is marked with a different color and each circle represents a given AOI. 
}
\label{fig:sample_optimization}
\end{figure*}

To better illustrate the optimization performance, Fig. \ref{fig:sample_optimization} presents a sample run considering all AOIs. The proposed method efficiently leverages the 3D city-scale map to optimize multiple ABS positions simultaneously while generating obstacle-avoiding routes (Fig. \ref{fig:ABS_routes}) for real-world deployment. Finally, Fig. \ref{fig:ABS_final_deployment} shows that, while all AOIs are covered, the remaining ABSs distribute in a lattice-like formation, useful for serving as relay nodes or providing additional coverage.

\subsubsection{Results for Orient. and Pow. Optimization (Method 1)}
Starting from the location configuration obtained in the experiment illustrated in Fig. \ref{fig:ABS_final_deployment}, we now apply the orientation optimization strategy described in Sec. \ref{sec:ABSorientation_overall} to showcase the performance of proposed optimization approach for improving ABSs' mutual interference when considering the whole urban site map. For this experiment, we consider ABS mounting directional antennas specified in the 5G NR standard TR 38.901 \cite{TR38901} and UE mounting half-wave dipole antennas, both with dual polarization (i.e., vertical and horizontal) and operating at a center frequency $f_c = 3.5$ GHz. For every coverage map computation, we consider map cells size of $10m \times 10m$ and downlink transmissions by setting the initial transmission power $P_i^{\text{tx}}$ in Watts for each ABS  to $43.0$ dBm. We consider Line-of-Sight (LOS), specular reflection, and diffuse scattering as possible rays interactions when generating the coverage maps during optimization and for coverage map plots. For each map, we consider $5$M rays equally distributed across the ABSs and up to $3$ ray interactions before reaching the destinations. For simplicity, we assign all objects and surfaces in the map with  \textit{concrete} material scattering properties, as defined by ITU-R P.2040 \cite{ITUR_P2040_3}. For these experiments, we use the RMSProp optimizer with learning rate $l_r = 0.1$ and $150$ training epochs, with early stopping criterion using $l_r$ decay of $0.5$ and patience of $5$ training epochs. For the loss hyper-parameters, we choose $\beta_{\text{L}} = 1.0$ for the smooth minimum computation and $\xi = 0.25$ as a scaling factor for the global average SIR loss term.

{\reviewI To evaluate the benefits of the proposed approach, Table \ref{tab:orient_optim_comparison} compares the initial effective SIR per ABS with the results from (i) the Max-Min approach ($\mathcal{L}_o$), (ii) a naïve average SIR maximization ($\mathcal{L}_{\text{avgSIR}}$), and (iii) uniform random parameter selection. While approach (ii) achieves the highest average SIR (5.76 dB), it creates significant performance inequality with a fairness index\footnote{Jain's Fairness Index: $FI = \frac{(\sum_{i=1}^{n} x_i)^2}{n \sum_{i=1}^{n} x_i^2}$, where $x_i$ are linear power values converted from dB, and $0 \leq FI \leq 1$ with higher values indicating more equitable distribution.} of only 0.115 and degrades worst-case performance (minimum SIR of -12.96 dB). This approach over-optimizes ABSs with already high SIR while severely penalizing those with lower values (e.g., ABS 1 and 4 at the cluster center). In contrast, the Max-Min approach (i) achieves superior fairness (fairness index of 0.306) and dramatically improves worst-case performance (minimum SIR of -1.35 dB vs -12.96 dB), demonstrating effective interference management by boosting lower-SIR ABSs (ABS 1, 2, 3, 4, and 6) by up to 8.24 dB while maintaining reasonable average performance. Random parameter selection, averaged over 50 samples, shows the poorest overall performance with negative average SIR (-0.65 dB) and very low fairness (0.178), validating the necessity of intelligent optimization. Table \ref{tab:orient_m1vm2_comparison} details the final orientations and transmit power settings for the Max-Min approach.}

\begin{table}[t]
\centering
\begin{tabular}{|l|l|l|l|l|}
\hline
SIR & \textbf{Initial} & \textbf{After} - $\mathcal{L}_o$ & \textbf{After} - $\mathcal{L}_{\text{avgSIR}}$ & \textbf{Random} \\ \hline
\textbf{ABS 1} & -9.595 & -1.352 & -12.963 & -12.773 \\ \hline
\textbf{ABS 2} & -0.189 & 1.421 & 7.941 & -3.246 \\ \hline
\textbf{ABS 3} & -1.537 & 1.214 & -3.302 & -3.636 \\ \hline
\textbf{ABS 4} & -4.213 & -0.542 & -10.621 & -10.530 \\ \hline
\textbf{ABS 5} & 4.625 & 0.471 & 8.950 & -2.795 \\ \hline
\textbf{ABS 6} & -1.165 & -0.439 & 4.660 & -7.507 \\ \hline
\textbf{ABS 7} & 19.760 & 11.057 & 19.729 & 14.660 \\ \hline
\textbf{ABS 8} & 25.527 & 12.728 & 32.018 & 18.893 \\ \hline
\textbf{ABS 9} & 5.666 & 0.454 & 6.422 & 0.981 \\ \hline
\textbf{ABS 10} & 3.248 & 0.192 & 4.793 & -0.519 \\ \hline \hline
\textbf{Average} & 4.213 & 2.520 & \textbf{5.763} & -0.647 \\ \hline
\textbf{Std. Dev.} & 10.209 & \textbf{4.765} & 12.662 & 9.648 \\ \hline
\textbf{Minimum} & -9.595 & \textbf{-1.352} & -12.963 & -12.773 \\ \hline
\textbf{Jain's Fairness} & 0.157 & \textbf{0.306} & 0.115 & 0.178 \\ \hline

\end{tabular}
\caption{{\reviewI Effective SIR (dB) experienced by each ABS before and after optimization. The $\mathcal{L}_o$ approach achieves superior worst-case performance (Min SIR) and fairness (Std. Dev., Jain's Fairness) compared to naïve average SIR maximization, demonstrating its effectiveness in improving poorly-performing ABSs while maintaining overall system balance.} }
\label{tab:orient_optim_comparison}
\end{table}


\subsubsection{Results for Orient. and Pow. Optimization (Method 2)}
\label{sec:perf_Lw}

While the previous strategy optimizes mutual interference across the map, we now evaluate the approach from Sec. \ref{sec:ABSorientation_AOI} to enhance SIR in specific AOIs. This method reuses the Sionna coverage map and optimization parameters, adjusting only the initial learning rate ($l_r = 0.05$) and setting the smooth minimum weighting temperature to $T=25$.

\begin{figure*}[t]
\centering
\includegraphics[width=0.95\textwidth]{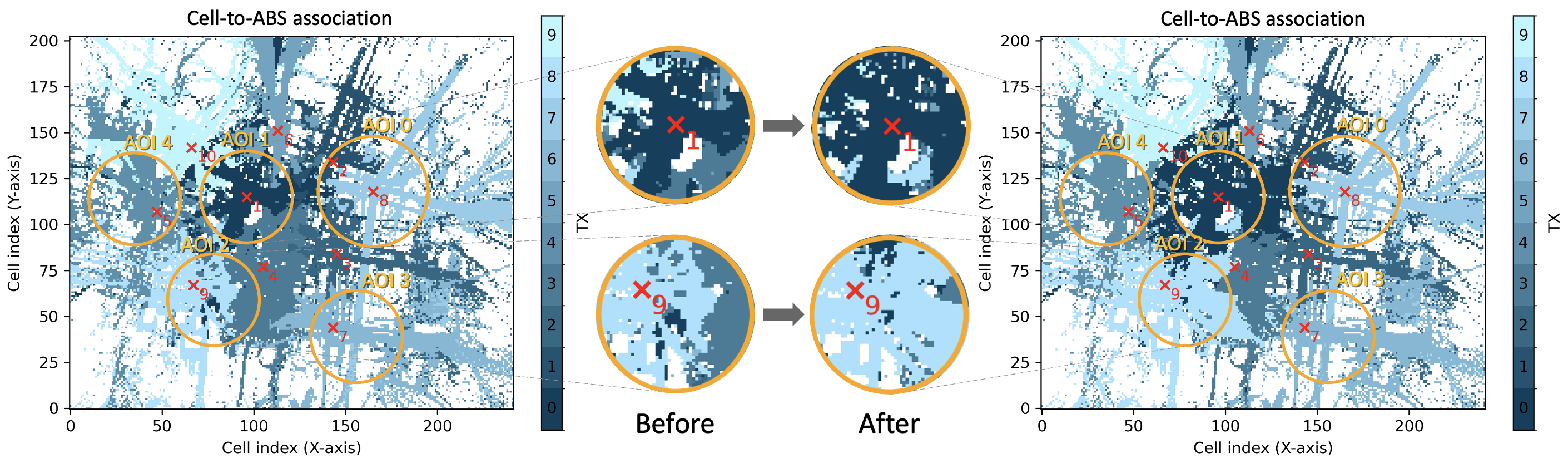}
\caption{{\reviewI  Cell-to-ABS association showing SIR-based coverage dominance before (left) and after (right) orientation and power optimization using the $L_w$ approach for sample scenario in Fig. \ref{fig:ABS_final_deployment}. Each cell on the ground surface is colored according to the ABS experiencing the highest SIR at that location, demonstrating improved coverage allocation within AOIs and more defined service boundaries. Quantitative SIR improvements are detailed in Table~\ref{tab:orient_SIR_difference} (up to +12.98 dB gains) and validated through AODT time-series average measurements in Figure~\ref{fig:AODT_valid_AOI} (3-20 dB improvements). Highlighted regions show AOIs with the most significant optimization impact.}}
\label{fig:orient_optim_method2}
\end{figure*}

{\reviewI  To evaluate the benefits of the proposed approach, we analyze the cell-to-ABS association patterns before and after optimization with $\mathcal{L}_w$ method. Figure~\ref{fig:orient_optim_method2} presents association maps where each cell is colored according to the ABS that provides the \textbf{highest SIR} (i.e., \textit{serving} ABS) at that location. The association maps reveal several key improvements after optimization: (i) coverage regions become more homogeneous within each AOI, ensuring UEs in critical areas experience consistent service from their designated serving ABS; (ii) boundaries between different ABS coverage zones become more clearly defined, reducing potential handover instabilities; and (iii) ABSs outside AOIs appropriately adjust their radiation patterns to minimize interference within AOIs. 

Table \ref{tab:orient_SIR_difference} quantifies the SIR improvements achieved by the weighted AOI optimization ($\mathcal{L}_w$) for the sample scenario, showing substantial gains for serving ABSs in most AOIs (e.g., +11.51 dB for ABS 1 in AOI 1, +9.57 dB for ABS 9 in AOI 2, and +5.17 dB for ABS 5 in AOI 4) and degradation of less than 0.5 dB for AOI 0 and AOI 3. Notably, the random baseline consistently degrades performance on average across all AOIs, with serving ABSs experiencing SIR losses ranging from -6.14 dB to -10.56 dB. This stark contrast demonstrates improvements of 8.18-17.65 dB achieved by our optimization over random parameter selection, with particularly strong gains in AOIs 1 and 2 (+17.65 dB and +15.73 dB respectively). Table \ref{tab:orient_SIR_difference_multi_deployment} extends this analysis across 50 different ABS deployments, confirming the robustness of the proposed approach. The average results show consistent positive gains for most AOIs (+12.98 dB, +10.01 dB, and +1.08 dB for AOIs 1, 2, and 4 respectively), with only modest reductions in AOIs 0 and 3 (-1.21 dB and -1.70 dB). Note that in some configurations the SIR gains might be lower due to multiple ABSs positioned within the same AOI, although the average values confirm that proposed approach is effective in keeping intra-AOI interference at minimal levels. These results validate that accurate power and orientation control are essential for efficient SIR management in autonomous ABS deployment, as random parameter selection leads to systematic performance degradation. The final configurations for the sample scenario are detailed in Table \ref{tab:orient_m1vm2_comparison} for reproducibility, highlighting how serving ABSs are assigned higher power levels and fine-tuned orientations to enhance effective SIR in their respective AOIs.}

{\reviewI 
\subsection{Computational Performance Evaluation For Practical Deployment}
We note that real-time operation analysis is beyond the scope of this paper, which focuses on demonstrating the feasibility and effectiveness of multi-digital twin optimization frameworks for ABS deployment. Our empirical measurements reveal distinct performance characteristics for the two optimization phases:
\begin{itemize}
    \item \textbf{Location Optimization:} Achieves $0.0371$s per iteration using only TensorFlow compiled operations and geometric calculations (without differentiable ray tracing). For the experimental scenario with $5$ AOIs and $10$ ABSs, complete optimization converges in approximately $92.75$s with a maximum of 2500 iterations, though early stopping significantly reduces this in practice. The computational efficiency stems from our gradient-based approach using only geometric loss functions.
    \item 
\textbf{Orientation and Power Optimization:} Gradient optimization via differentiable ray tracing (SionnaRT) incurs substantially higher computational costs. Each training iteration requires approximately $15$s in our experimental settings, even with GPU acceleration. The computational complexity scales with map size, number of ABSs, polygon count, ray count, and ray interaction limits.
Nevertheless, this performance profile is well-suited for practical ABS deployment scenarios where:
\begin{enumerate}
\item \textbf{Mission planning phase:} Combined optimization time (typically 2-5 minutes) is acceptable for pre-mission deployment planning;
\item \textbf{Adaptive repositioning:} The algorithm can provide updated configurations during flight operations, as typical ABS battery life is 20-100 minutes~\cite{top10drones};
\item \textbf{Hierarchical deployment:} Fast location optimization provides initial positioning and initial coverage, with orientation refinement performed as needed during real-world operations using the Digital Twins;
\item \textbf{Scalability:} Optimization time scales with scenario complexity, enabling faster solutions for smaller deployments.
\end{enumerate}
\end{itemize}

Future performance improvements could be achieved through careful hyperparameter tuning and leveraging improved SionnaRT implementations, built solely on Mitsuba3/Dr.Jit instead of the current\footnote{Tested with Sionna 0.19.2 and Mitsuba 3.6.4} Mitsuba+TensorFlow architecture and that allows sampling techniques such as Russian Roulette (RR) for more efficient Ray Tracing differentiation\cite{sionna_technical_report}.
While real-time optimization presents interesting challenges for future work, the current framework demonstrates the foundational capability of multi-DT systems for autonomous ABS deployment optimization.
\subsubsection{Profiling Analysis}
Profiling analysis using NVIDIA Nsight Systems revealed that the computational pipeline exhibits a nearly balanced distribution between GPU kernel execution (51.6\%) and memory operations (48.4\%), indicating that the application operates in a memory-bound regime where data movement overhead significantly impacts overall performance. This memory-to-computation ratio suggests that while the GPU cores are effectively utilized for mathematical operations, substantial execution time is consumed by host-device data transfers and GPU memory management operations.
The predominance of memory operations presents both a performance bottleneck and an optimization opportunity, as memory transfer patterns are often more amenable to algorithmic improvements than pure computational limitations. Such a profile typically indicates potential for performance gains through strategies that minimize data movement, such as maintaining GPU-resident data structures across multiple computational phases, implementing asynchronous memory transfers overlapped with kernel execution, and consolidating operations to reduce the frequency of host-device communication. The near-equal distribution between computation and memory operations underscores the importance of considering data locality and transfer efficiency as primary factors in the optimization strategy, rather than focusing solely on computational algorithm improvements.
}

{\reviewII 
\subsubsection{Gradient-Based vs. Learning-Based Approach Rationale}

Our gradient-based optimization approach is specifically designed for systems with complete environmental information obtained via Digital Twin simulations. As mentioned in Sec.~\ref{sec:relatedwork}, unlike RL approaches that require extensive exploration and training, our method leverages the differentiable nature of the targeted Multi-DT system for direct optimization of wireless system parameters. RL-based approaches would introduce unnecessary computational overhead including data collection phases, model architecture exploration and training, and deployment complexity. Since our proposed Multi-DT framework provides complete environmental knowledge with differentiable propagation models, direct gradient optimization offers superior efficiency with immediate applicability to new scenarios without retraining requirements.

The 0.0371s per iteration performance for location optimization and deterministic convergence properties demonstrate the practical advantages of leveraging complete environmental information over trial-and-error learning approaches.
}



\begin{table}[t]
\centering
\begin{tabular}{|l|lllll|}
\hline
\multicolumn{6}{|c|}{Average AOI Effective SIR (dB) difference} \\ \hline \hline
\multicolumn{6}{|c|}{Weighted AOI SIR optimization - $\mathcal{L}_w$ (sample scenario)} \\ \hline
\multicolumn{1}{|l|}{} & \multicolumn{1}{l|}{\textbf{AOI 0}} & \multicolumn{1}{l|}{\textbf{AOI 1}} & \multicolumn{1}{l|}{\textbf{AOI 2}} & \multicolumn{1}{l|}{\textbf{AOI 3}} & \textbf{AOI 4} \\ \hline
\multicolumn{1}{|l|}{\textbf{ABS 1}}  & \multicolumn{1}{l|}{+0.99}  & \multicolumn{1}{l|}{\textbf{+11.51}} & \multicolumn{1}{l|}{-0.99}  & \multicolumn{1}{l|}{+0.05}   & +0.23  \\ \hline
\multicolumn{1}{|l|}{\textbf{ABS 2}}  & \multicolumn{1}{l|}{-0.02} & \multicolumn{1}{l|}{-4.51}  & \multicolumn{1}{l|}{-2.27}  & \multicolumn{1}{l|}{-0.09}  & -0.25 \\ \hline
\multicolumn{1}{|l|}{\textbf{ABS 3}}  & \multicolumn{1}{l|}{-0.79} & \multicolumn{1}{l|}{-4.94}  & \multicolumn{1}{l|}{-6.22}  & \multicolumn{1}{l|}{-0.77}  & -0.26 \\ \hline
\multicolumn{1}{|l|}{\textbf{ABS 4}}  & \multicolumn{1}{l|}{0.01}  & \multicolumn{1}{l|}{-10.37} & \multicolumn{1}{l|}{-8.67}  & \multicolumn{1}{l|}{+2.48}   & -0.48 \\ \hline
\multicolumn{1}{|l|}{\textbf{ABS 5}}  & \multicolumn{1}{l|}{+0.01}  & \multicolumn{1}{l|}{-1.40}  & \multicolumn{1}{l|}{-4.89}  & \multicolumn{1}{l|}{+0.07}   & \textbf{+5.17} \\ \hline
\multicolumn{1}{|l|}{\textbf{ABS 6}}  & \multicolumn{1}{l|}{+0.08}  & \multicolumn{1}{l|}{-5.45}  & \multicolumn{1}{l|}{-1.70}  & \multicolumn{1}{l|}{-0.004}  & -0.93 \\ \hline
\multicolumn{1}{|l|}{\textbf{ABS 7}}  & \multicolumn{1}{l|}{-0.51} & \multicolumn{1}{l|}{-2.33}  & \multicolumn{1}{l|}{-5.08}  & \multicolumn{1}{l|}{\textbf{-0.25}}  & +0.02  \\ \hline
\multicolumn{1}{|l|}{\textbf{ABS 8}}  & \multicolumn{1}{l|}{\textbf{-0.49}} & \multicolumn{1}{l|}{+0.28}   & \multicolumn{1}{l|}{-1.12}  & \multicolumn{1}{l|}{-0.62}  & +0.01  \\ \hline
\multicolumn{1}{|l|}{\textbf{ABS 9}}  & \multicolumn{1}{l|}{+0.03}  & \multicolumn{1}{l|}{+0.11}   & \multicolumn{1}{l|}{\textbf{+9.57}}   & \multicolumn{1}{l|}{+1.94}   & -1.90 \\ \hline
\multicolumn{1}{|l|}{\textbf{ABS 10}} & \multicolumn{1}{l|}{+0.01}  & \multicolumn{1}{l|}{-5.64}  & \multicolumn{1}{l|}{-2.90}  & \multicolumn{1}{l|}{+0.0003}   & -5.79 \\ \hline \hline
\multicolumn{6}{|c|}{Serving ABS only - Random Baseline (sample scenario, 50 runs mean)} \\ \hline
\multicolumn{1}{|l|}{} & \multicolumn{1}{l|}{\textbf{AOI 0}} & \multicolumn{1}{l|}{\textbf{AOI 1}} & \multicolumn{1}{l|}{\textbf{AOI 2}} & \multicolumn{1}{l|}{\textbf{AOI 3}} & \textbf{AOI 4} \\ \hline
\multicolumn{1}{|l|}{\textbf{ABS (serving)}}  & \multicolumn{1}{l|}{-8.92}  & \multicolumn{1}{l|}{-6.14} & \multicolumn{1}{l|}{-8.43}  & \multicolumn{1}{l|}{-10.17}   & -10.56  \\ \hline 
\end{tabular}

\caption{{\reviewI Difference of average SIR (dB) for each AOI after orientation and power optimization. ABSs serving a given AOI (i.e., the one with highest SIR) are indicated in bold. Results are reported for sample scenario in Fig. \ref{fig:ABS_final_deployment}.}}
\label{tab:orient_SIR_difference}
\end{table}

\begin{table}[t]
\centering
\begin{tabular}{|l|lllll|}
\hline
\multicolumn{6}{|c|}{Average AOI Effective SIR (dB) difference} \\ \hline \hline
\multicolumn{6}{|c|}{Serving ABS only - $\mathcal{L}_w$ (50 scenarios/deployments mean) } \\ \hline
\multicolumn{1}{|l|}{} & \multicolumn{1}{l|}{\textbf{AOI 0}} & \multicolumn{1}{l|}{\textbf{AOI 1}} & \multicolumn{1}{l|}{\textbf{AOI 2}} & \multicolumn{1}{l|}{\textbf{AOI 3}} & \textbf{AOI 4} \\ \hline
\multicolumn{1}{|l|}{\textbf{ABS (serving)}}  & \multicolumn{1}{l|}{-1.21}  & \multicolumn{1}{l|}{+12.98} & \multicolumn{1}{l|}{+10.01}  & \multicolumn{1}{l|}{-1.70}   & +1.08  \\ \hline
\end{tabular}
\caption{{\reviewI Average SIR difference for each AOI under 50 different ABSs deployments.}}
\label{tab:orient_SIR_difference_multi_deployment}
\end{table}

\begin{table}[t]
\centering
\resizebox{\columnwidth}{!}{
\begin{tabular}{|l|l|l|l|l|l|l|l|l|}
\hline
\textbf{ABS} & \multicolumn{2}{l|}{\textbf{Coordinates}} & \multicolumn{3}{l|}{\textbf{Orient. \& Power} - $\mathcal{L}_o$} & \multicolumn{3}{l|}{\textbf{Orient. \& Power} - $\mathcal{L}_w$} \\ \hline
 & \multicolumn{1}{l|}{X (m)} & Y (m) & \multicolumn{1}{l|}{$\phi$} & \multicolumn{1}{l|}{$\theta$} & $P^{\text{tx}}$ & \multicolumn{1}{l|}{$\phi$ } & \multicolumn{1}{l|}{$\theta$} & $P^{\text{tx}}$ \\ \hline
\textbf{1}  & \multicolumn{1}{l|}{-246.60} & 144.50 & 17.72  & 154.29 & 42.96  & -30.07 & 88.96  & 43.00  \\ \hline
\textbf{2}  & \multicolumn{1}{l|}{223.43}  & 334.20 & -7.48  & 25.71  & 40.78  & 37.09  & 63.33  & 37.11  \\ \hline
\textbf{3}  & \multicolumn{1}{l|}{245.27}  & -174.25 & -25.37 & 25.71  & 39.51  & -4.80  & 77.19  & 35.23  \\ \hline
\textbf{4}  & \multicolumn{1}{l|}{-148.59} & -243.15 & 2.54   & 120.73 & 41.88  & -42.43 & 25.71  & 35.30  \\ \hline
\textbf{5}  & \multicolumn{1}{l|}{-735.17} & 56.92  & 81.42  & 104.03 & 38.91  & -28.43 & 138.64 & 39.49  \\ \hline
\textbf{6}  & \multicolumn{1}{l|}{-68.51}  & 503.55 & 11.42  & 25.71  & 39.75  & 38.94  & 74.68  & 35.08  \\ \hline
\textbf{7}  & \multicolumn{1}{l|}{225.58}  & -574.39 & -60.03 & 25.71  & 32.10  & -41.10 & 56.94  & 39.03  \\ \hline
\textbf{8}  & \multicolumn{1}{l|}{450.19}  & 168.14 & -0.84  & 39.74  & 26.89  & -1.79  & 81.77  & 40.42  \\ \hline
\textbf{9}  & \multicolumn{1}{l|}{-535.52} & -343.17 & -68.07 & 134.79 & 34.60  & 114.99 & 133.16 & 42.26  \\ \hline
\textbf{10} & \multicolumn{1}{l|}{-545.98} & 409.48 & -99.82 & 154.29 & 37.14  & -70.22 & 126.54 & 38.13  \\ \hline
\end{tabular}
}
\caption{Configurations obtained with proposed optimization frameworks for sample scenario (Fig. \ref{fig:ABS_final_deployment}). $\phi$ is the azimuth angle of rotation in degrees, $\theta$ is the mechanical tilt in degree, $P^{\text{tx}}$ is the transmission power in dBm.}
\label{tab:orient_m1vm2_comparison}
\end{table}

\section{{\reviewI  Cross-}Validation of ABS Deployment in AODT}
\label{sec:AODT_validation}

\begin{figure}[t]
\centering
\includegraphics[width=1.\columnwidth]{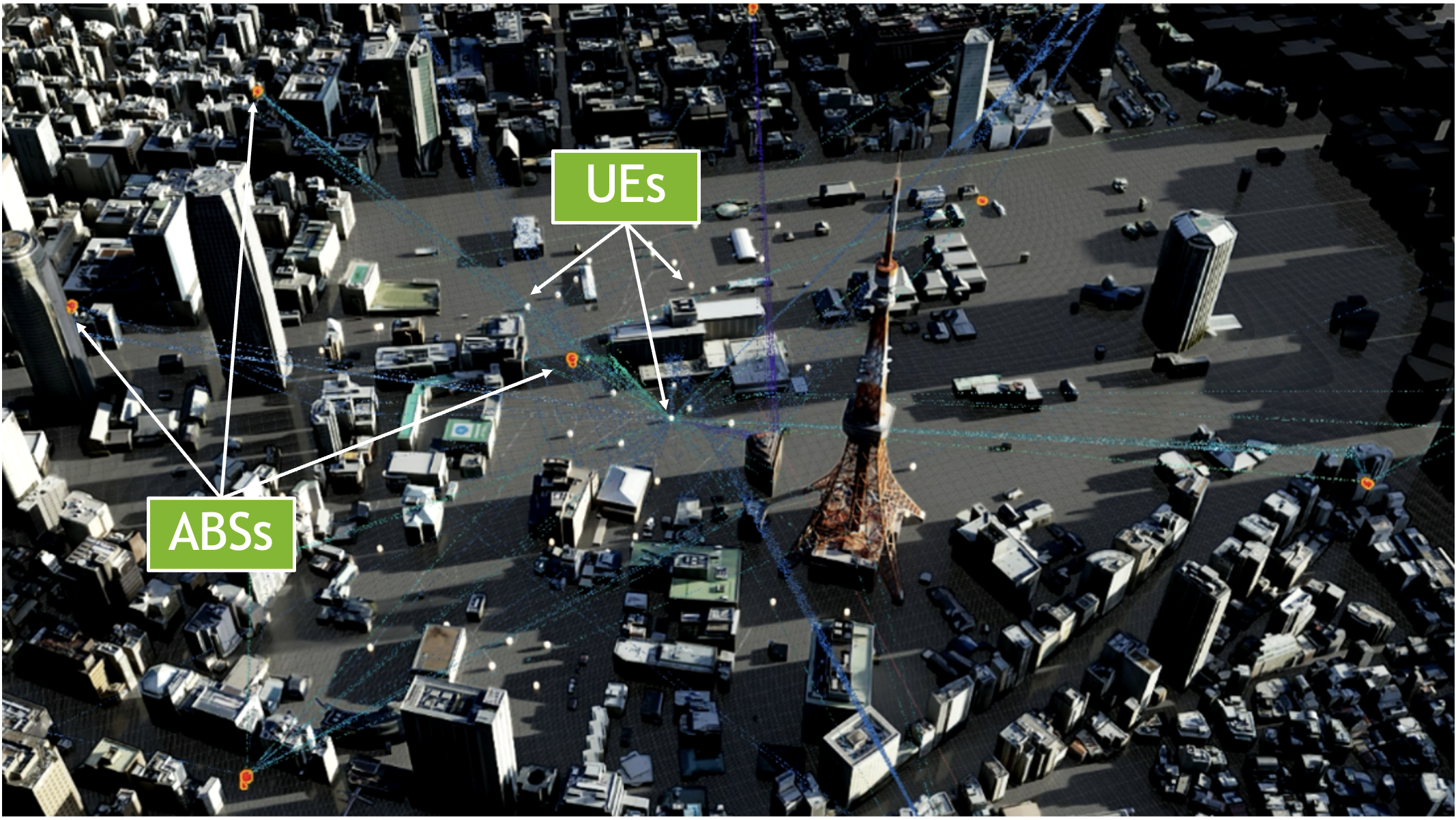}
\caption{Live capture of ABS configuration imported from Sionna to AODT. UEs placed in AOI 1 and rays shown for all ABSs and a sample UE.}
\label{fig:AODT_live_screenshot}
\end{figure}

In this section, we validate ABS deployments obtained via Sionna optmization by measuring UE-perceived signal strength over time using AODT-generated Channel Impulse Responses (CIRs) for point-to-point communications. Each simulation runs for $60$s at a granularity of $1$s time steps, leveraging AODT’s procedural UE generation. We focus on validating the $\mathcal{L}_w$ optimization approach for the scenario in Sec. \ref{sec:perf_Lw} (see Table \ref{tab:orient_m1vm2_comparison} for full parameters) by mapping each AOI to a \textit{spawn zone} in AODT. Due to AODT 1.1.1 constraints, we define square spawn zones centered on each AOI with edge lengths $2 r_m$ and collect measurements for a single AOI at a time. Each simulation deploys $U=50$ UEs moving within their AOI.  We configure AODT with simulation parameters matching Sionna (where applicable), i.e. $f_c = 3.5$ GHz, ITU concrete material for surfaces, TR 38.901\footnote{Imported into AODT via custom CSV descriptor.} antenna pattern for ABSs, halfwave dipole for UEs, and $500$K rays per ABS. To optimize efficiency, we assume only vertical polarization for ABS and UE antennas. ABS parameters (location, orientation, power) are exported from Sionna in JSON format, then imported into our modified AODT code, which also deploys spawn zones per AOI. UEs are initialized with fixed mechanical azimuth and tilt $\phi = \theta = 0.0$. We conduct multi-UE simulations for each AOI, separately collecting CIR for downlink transmissions and location data for all ABSs and UEs at each simulation step, stored in an AODT database. Fig. \ref{fig:AODT_live_screenshot} shows a screenshot of the imported configuration during a live simulation.  


We compute the signal strength perceived between the $u$-th UE and $i$-th ABS at time step $t$ as the sum the channel gains for every valid path\footnote{As of AODT 1.1.1, CIR data stored for each antenna pair is limited to the strongest 500 channel taps.} multiplied by the ABS's transmission power computed during Sionna optimization phase:  $P_{t,i,u}^{\text{rx}} = P_i^{\text{tx}} \sum_{r=1}^{N_r} |h_r|^2$, where $h_r \in \mathbb{C}$ is the complex channel gain for the $r$-th channel tap (or ray path) and $N_r$ is the total number of valid paths. In order to compute the SIR based on the power perceived from all other ABSs in the map, we compute the total signal power $P_{t,u}^{\text{tot}} = \sum_{j=1, j \neq i}^N P_{t,j,u}^{\text{rx}}$
and then compute the SIR for each $u$-th UE given an ABS $i$ and simulation time $t$ as $\text{SIR}_{t,u} = \frac{P_{t,i,u}^{\text{rx}}}{P_{t,u}^{\text{tot}}}$.

\begin{figure}[t]
\centering
\includegraphics[width=\figurewidthI]{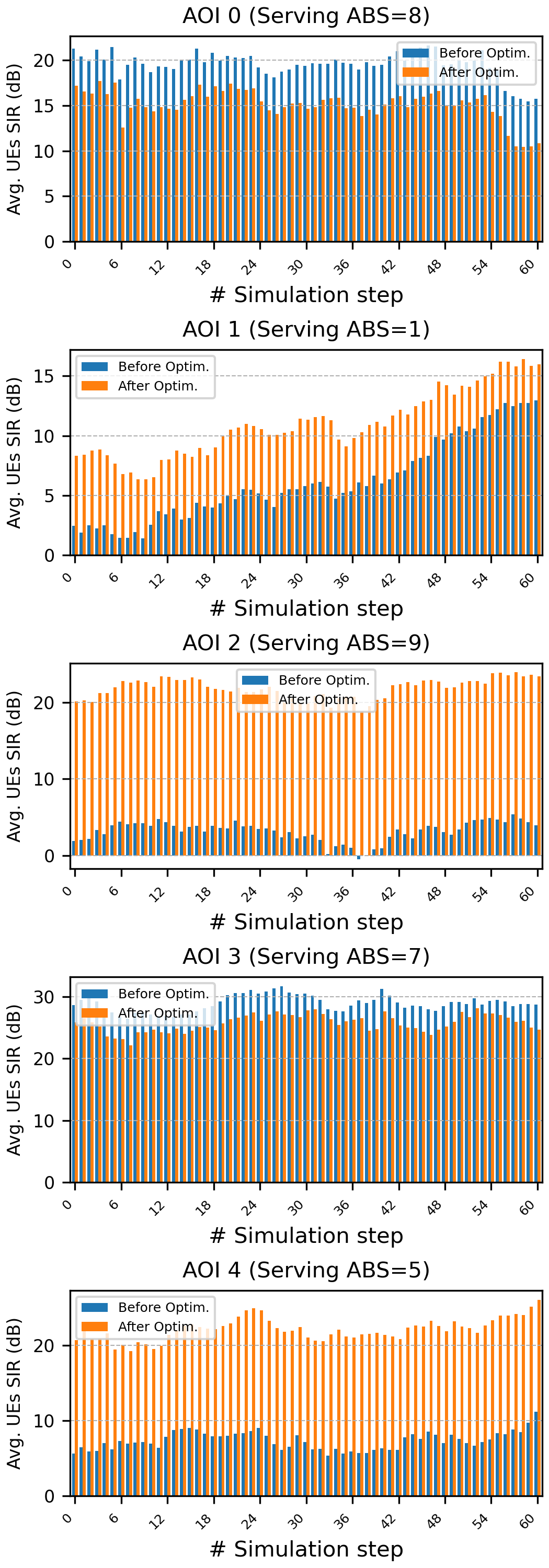}
\caption{SIR measured with AODT, averaged over 50 UEs moving within a given AOI, using ABSs' parameters obtained before and after orientation and power optimization via Sionna. }
\label{fig:AODT_valid_AOI}
\end{figure}

\begin{figure}[t]
\centering
\includegraphics[width=\figurewidthII]{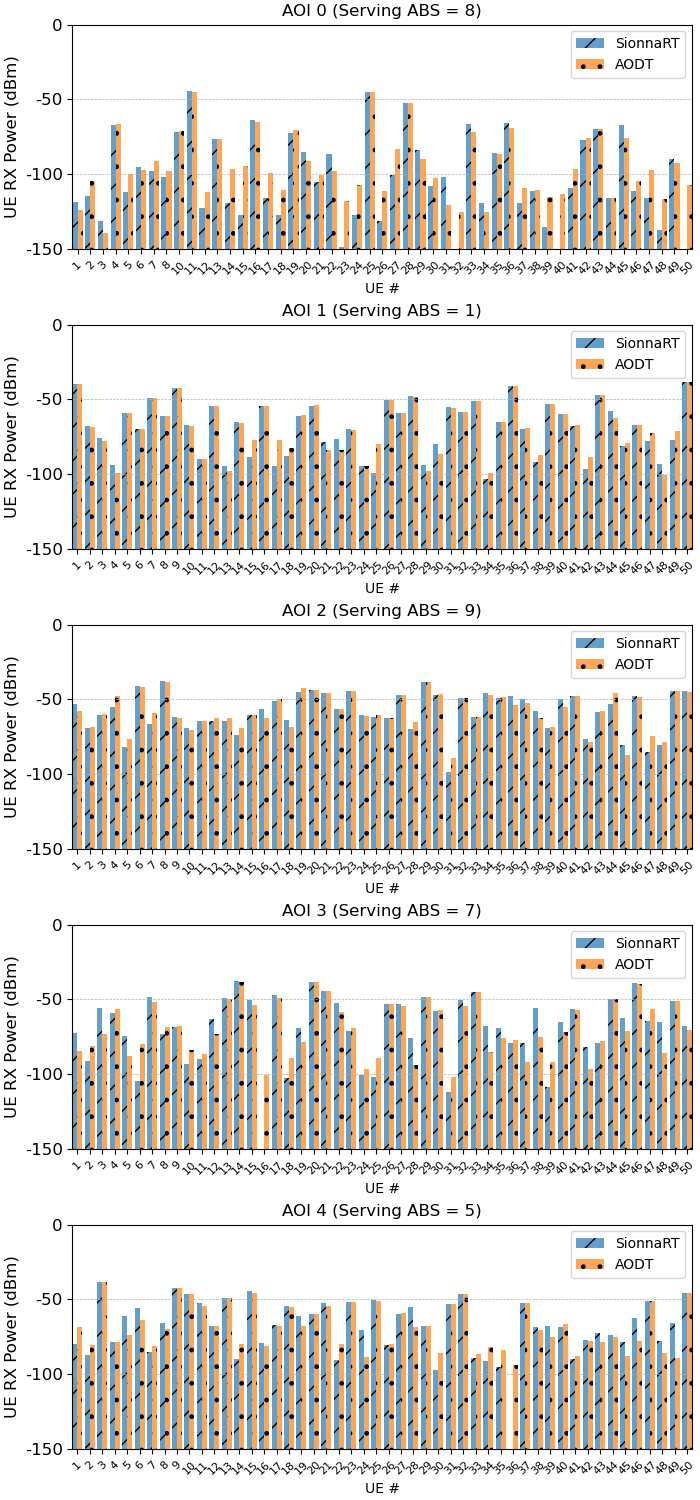}
\caption{Comparison between receiver power measured at each UE in AODT and Sionna at Simulation Frame $0$, using TR 38.901 antenna pattern at Transmitter (ABS) and halfwave dipole at Receiver (UE).}
\label{fig:AODT_vs_Sionna_measures}
\end{figure}

Fig. \ref{fig:AODT_valid_AOI} presents SIR measurements for each AOI and mobile user in AODT. The SIR values, averaged over 50 UEs per AOI, are computed using the serving ABS's reference signal power, both before and after orientation and power optimization obtained via Sionna. The optimized parameters improve SIR in AOIs 1, 2, and 4, with gains between 3 and 20 dB. Conversely, AOIs 0 and 3, which already had high SIR, experience slight reductions, due to adjustments favoring AOIs with poorer SIR. However, their SIR remains sufficient for reliable communication, demonstrating how the framework's provides solutions generalizable across multiple DTs considering real-world scenarios. Finally, Fig. \ref{fig:AODT_vs_Sionna_measures} compares UEs' channel gains from both DTs using identical simulation parameters (where applicable) for the first simulation frame, showing consistency between the tools and further validating the feasibility of the proposed approach.

\section{ABS Deployment Adaptation For Mission-Critical Scenarios}
\label{sec:ABSadaptation}
We have seen how Sionna can be used to perform deployment optimizations for wireless devices to produce configurations validated via AODT. In this section, 
we explore a use case that reverse the data flow between the two DTNs to address mission-critical UEs' severe coverage loss due to shadowing. We propose a threshold-based algorithm to detect signal drops in AODT simulations and adapt trajectory optimization methods via Sionna to adjust ABS parameters, improving UE signal power. Once critical UE paths are identified, they can be used to predict future issues, enabling DTNs to provide pre-computed recovery solutions or serve as training data for learning-based models.

\subsection{Signal Drop Detection}
Assuming a single critical UE $u$ and its associated ABS $i$, we compute the receive power $P_{t,i,u}^{\text{rx}}$ expressed in Watts (see Sec. \ref{sec:AODT_validation}) for every $t$-th simulation step, obtaining a power measurements array $\boldsymbol{p}_{i,u} = \left[ P_{1,i,u}^{\text{rx}}, ..., P_{N_t,i,u}^{\text{rx}} \right]$, where $N_t$ is the number of simulated steps. Using this information, we want to identify one or more sudden drops in signal coverage for any UE $u$ that span several consecutive simulation steps, indicating its passage in shadowed areas such as narrow streets between tall buildings or other blocking entities. To this extent, we propose a threshold-based detection algorithm that works as follows:

\begin{enumerate}
    \item \textbf{Detection}: Scans $\boldsymbol{p}_{i,u}$ for received power drops below a threshold $T_{min}$.
    \item \textbf{Confirmation}: A drop is confirmed only if $c_{min}$ consecutive measurements remain below $T_{min}$.
    \item \textbf{Spurious Peak Tolerance}: Allows up to $s_p$ temporary peaks above $T_{min}$ before ending detection.
    \item \textbf{Finalization}: If peaks exceed $s_p$, records start ($t_s$) and end ($t_e$) frames of the drop, computing its duration $T_d = t_e - t_s$.
    \item \textbf{Repeat}: Continues scanning $\boldsymbol{p}_{i,u}$ to detect further drops.
\end{enumerate}

By adjusting parameters $T_{min}$, $c_{min}$ and $s_p$, we can define different levels of tolerance for the power threshold and granularity of the detection algorithm, depending on the level of reliability required.


\subsection{ABS Recovery Trajectory and Configuration}
Once the power drops have been identified, for each drop interval we want to find a new configuration of the serving ABS that allows to improve the coverage conditions of critical UEs and provide more reliable communication. Once $(t_s, t_e)$ drop start and end frames are obtained, we extract from AODT the relative critical UE consecutive $T_d$ route positions in the map, indicated as $D_{\text{c}} = \left\{ \boldsymbol{d}_{t_s}, \boldsymbol{d}_{t_s + 1}, ..., \boldsymbol{d}_{t_e} \right\}$ where each $\boldsymbol{d}_t$ corresponds to the 2D $(x,y)$ coordinate of UE at time $t$. The proposed recovery approach aims to (i) find a trajectory via gradient-based optimization from the deployment location of the ABS toward the critical UE experiencing the power drop and (ii) compute the new ABS coordinates to be applied during simulation for the recovery operation. 

For the first goal, we first extract the UE's route coordinate corresponding to the \textit{middle point} of the drop interval, indicated as $\boldsymbol{d}_m$ at time $t_m$, assuming $t_s < t_m < t_e$. We then use $\boldsymbol{d}_m$ as destination point for the trajectory computation, based on the technique explained in Sec. \ref{sec:ABSlocation} and optimized via the following simplified criteria:
\begin{equation}
    \mathcal{L}_d = \left\|\boldsymbol{p}_i-\boldsymbol{c}_k\right\| + P_b
\end{equation}
which aims to minimize the distance between the current ABS location and $\boldsymbol{d}_m$ while avoiding collisions with buildings through the computation of $P_b$ penalty term. Once the destination is reached, we can employ the Sionna differentiable Ray Tacer to further refine the final recovery configuration, by optimizing its orientation and transmission power through one of the methods proposed in Sec. \ref{sec:ABS_orient_pow_optim}. 

Once the optimization converges towards the destination point and the trajectory point for each optimization iteration are obtained, we then compute the adjusted simulation coordinates by dividing the recovery operation in three distinct phases:
\begin{itemize}
    \item \textbf{\textit{Reaction} phase}: We extract $\lfloor T_d / 2 \rfloor$ equally distanced points from the trajectory obtained via the optimizer, corresponding to the locations that the ABS will have to follow while reacting to the signal drop detection in the $(t_s, t_m)$ simulation interval;
    \item \textbf{\textit{Stationary} phase}: in this phase we consider the ABS hovering over the critical UE route during the signal drop in the $(t_m, t_e)$ simulation steps, in order to improve the signal coverage for the remainder of the drop interval;
    \item \textbf{\textit{Return} phase}: Once the drop interval is concluded, the ABS is instructed to fly back to its original deployment location, until the next coverage loss is detected. To do so, it follows the same $\lfloor T_d / 2 \rfloor$ coordinates from \textit{Reaction} phase in reverse order.
\end{itemize}

\begin{figure}[t]
\centering
\includegraphics[width=0.8\columnwidth]{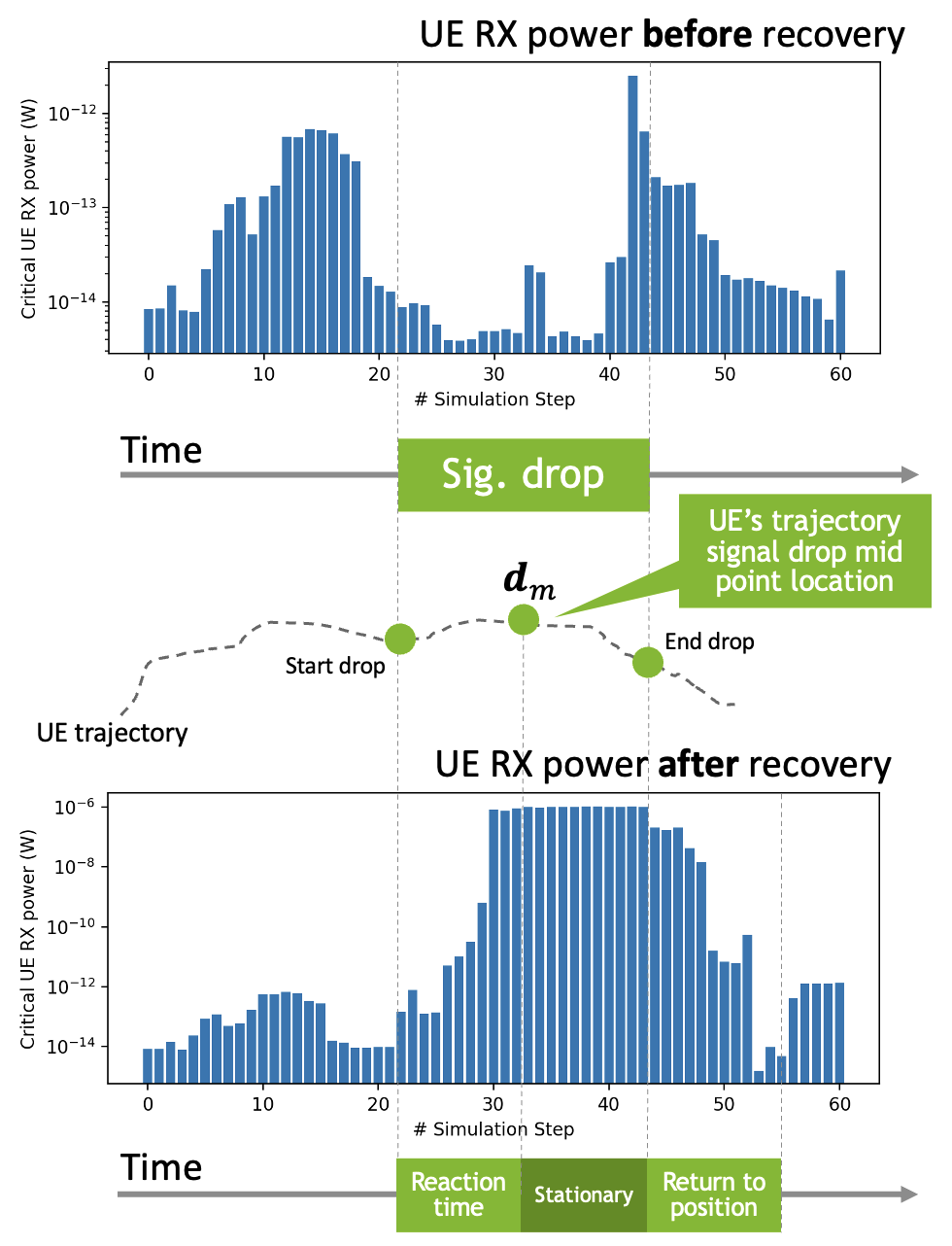}
\caption{Signal drop recovery mechanism overview for mission-critical UE. The bar plots show the receive power computed via AODT perceived by the UE experiencing a signal drop before (above) and after (below) the recovery mechanism is activated.}
\label{fig:sigdrop_recovery_schema}
\end{figure}

Fig. \ref{fig:sigdrop_recovery_schema} offers a visual overview of the recovery phases discussed above.

\begin{figure}[t]
\centering
\includegraphics[width=0.9\columnwidth]{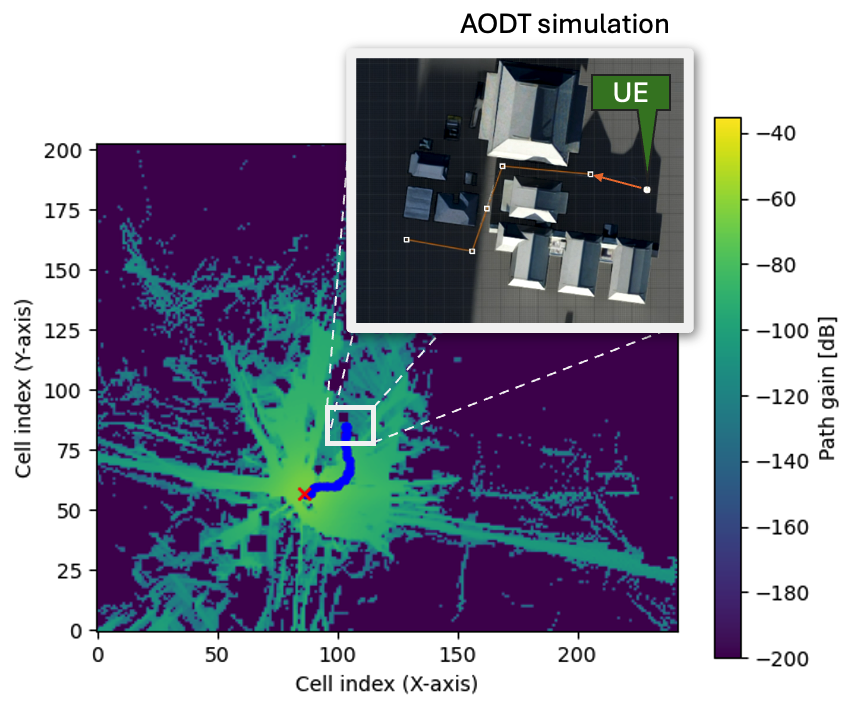}
\caption{Trajectory (blue) of ABS computed via gradient-descent for signal recovery of critical UE, projected over Sionna's coverage map of initial ABS position (marked with a red cross). In the cutout: Trajectory of critical UE simulated in AODT (moving right to left).}
\label{fig:sigdrop_recovery_ABS}
\end{figure}

\vspace{-1.5em}
\subsection{Experimental evaluation}

To evaluate the proposed coverage drop detection and recovery mechanism, we consider a single ABS and one critical UE. Fig. \ref{fig:sigdrop_recovery_ABS} illustrates the targeted sample scenario, having an UE’s passing through buildings that cause signal blockages and an ABS positioned $\sim$320m away and hovering behind a tall obstacle. Over 60 time steps, the UE moves along this path while collecting power measurements. For this experiment, we set $T_{\text{min}} = 10^{-14}$ W, $c_{\text{min}} = 3$, and $s_p = 5$, using Adam optimizer with $l_r = 6.0$, early stopping ($0.5$ decay, patience = 3). A signal drop is detected between $t_s = 22$ and $t_e = 44$, triggering the recovery trajectory optimizer. After assigning trajectory points for recovery, accounting for obstacles, we re-run the same UE simulation including provided ABS mobility pattern\footnote{Using AODT customized source code to support Radio Units (RU) mobility.} and collect new power measurements. Fig. \ref{fig:sigdrop_recovery_schema} visualizes the coverage throughout the simulation, indicating signal drop and recovery phases. The plots confirm up to $\sim6$ orders of magnitude power improvement post-optimization via the second DTN, proving the proposed approach effective in dynamically enhancing signal converage for that specific UE.




\section{Conclusion and Future Work}

{\reviewI  This work demonstrates the potential of using multiple Digital Twins (DTs) to perform complex cyber-physical simulations in the context of wireless communications. It focuses on optimizing and validating the deployment of Airborne Base Stations in urban settings, evaluating multiple goals and varying system conditions. Proposed solutions aim to highlight the flexibility of gradient-based optimization combined with Ray Tracing and detailed 3D geometries that can be potentially transferred to real-world environments. This paves the way for new kinds of studies for large-scale wireless systems, enabling innovative applications and solutions that combine the powerful simulation capabilities of modern DTNs and generate data fueling novel research.

{\reviewII  The gradient-based Multi-DT approach demonstrates clear advantages over alternative methods by leveraging complete environmental information and differentiable system components, avoiding the computational overhead and training requirements of learning-based approaches while achieving superior performance metrics.}
Future research directions focus on addressing computational efficiency and enhancing deployment intelligence. The orientation and power optimization overhead could be significantly reduced through improved SionnaRT implementations built solely on Mitsuba3/Dr.Jit~\cite{sionna_technical_report}, enabling advanced sampling techniques such as Russian roulette and reducing the current 15-second per iteration time to real-time levels. Integration with Channel Knowledge Map (CKM) techniques~\cite{yang2025channel,li2022channel} represents another promising avenue, where CKM-derived channel predictions could provide intelligent initialization for gradient-based optimization, reducing convergence time from minutes to seconds while enabling hybrid approaches that use CKM for coarse positioning and high-fidelity ray tracing for critical parameters. The convergence of enhanced ray tracing performance and intelligent prior knowledge integration through bi-directional CKM-Multi-DT information flow could create self-improving systems that continuously refine environmental knowledge while adapting to real-time changes. 
}
\label{sec:conclusion}

\bibliographystyle{IEEEtran}
\bibliography{sample-base2}
\begin{IEEEbiography}[{\includegraphics[width=1in,height=1.25in,clip,keepaspectratio]{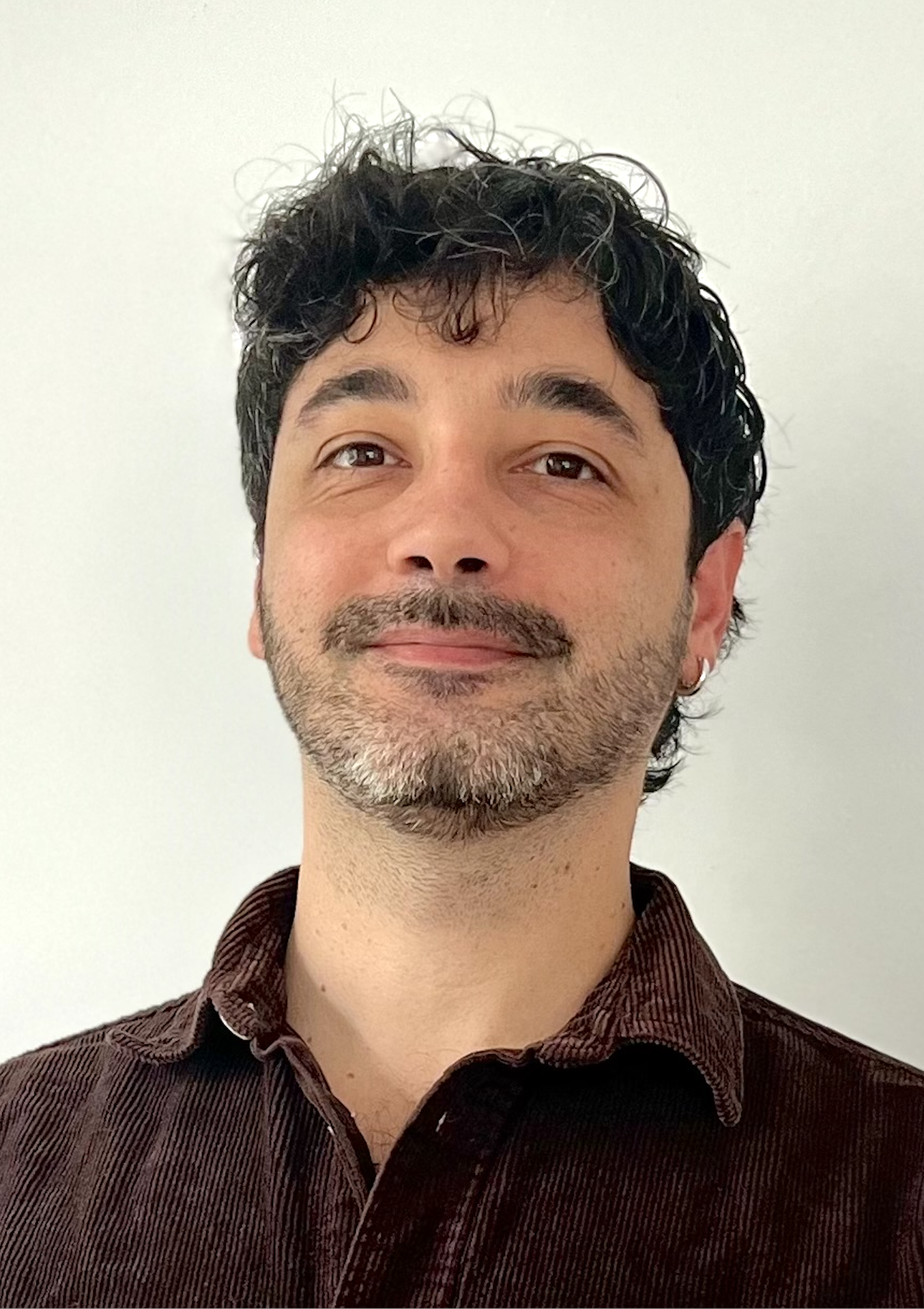}}]{Mauro Belgiovine}
earned his Ph.D. in Computer Engineering at Northeastern University (Boston, MA) in March 2025 under the guidance of Prof. Kaushik Chowdhury. He joined NVIDIA full-time in April 2025 to continue his research on deep learning applications to wireless communication, digital twins, semantic communications and generative AI.
\end{IEEEbiography}
\vspace{-3em}

\begin{IEEEbiography}[{\includegraphics[width=1in,height=1.25in,clip,keepaspectratio]{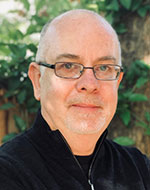}}]{Chris Dick} 
is a wireless architect with NVIDIA and the technical lead for the application of AI and machine learning to 5G and 6G wireless. In over 24 years working in signal processing and communications, he has delivered silicon and software products for 3G, 4G, and 5G baseband DSP and Docsis 3.1 cable access. He has performed research and delivered products for digital frontend (DFE) technology for cellular systems.
\end{IEEEbiography}
\vspace{-3em}

\begin{IEEEbiography}[{\includegraphics[width=1in,height=1.25in,clip,keepaspectratio]{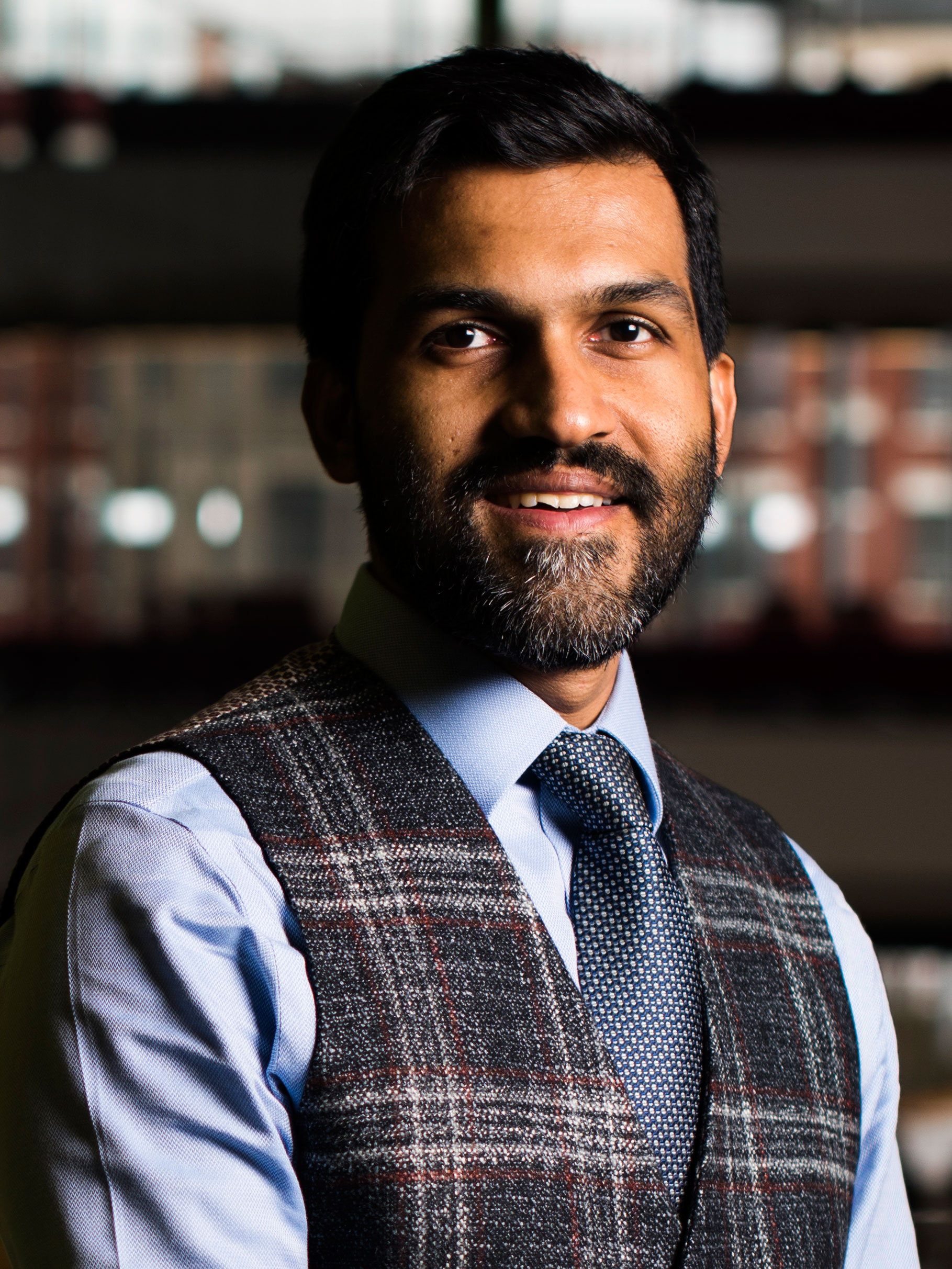}}]{Kaushik Chowdhury} 
is a Chandra Family Endowed Distinguished Professor in Electrical and Computer Engineering at The University of Texas at Austin. His research interests involve systems aspects of machine learning for agile spectrum sensing/access, unmanned autonomous systems, programmable and open cellular networks, and large scale experimental deployment of emerging wireless technologies.
\end{IEEEbiography}
\end{document}